\documentclass{aa}  
\usepackage{natbib}
\bibpunct{(}{)}{;}{a}{}{,} 
\usepackage{graphicx}

\usepackage{txfonts}

\begin{document} 

\title{Fitting isochrones to open cluster photometric data}

   \subtitle{III. Estimating metallicities from UBV photometry}

   \author{A. F. Oliveira\thanks{E-mail: adhimarflavio@unifei.edu.br}\inst{1}, H. Monteiro\inst{1}, W. S. Dias\inst{1} and
     T. C. Caetano\inst{1}\inst{2}} \institute{UNIFEI, IFQ -
     Instituto de F\'isica e Qu\'imica, Universidade Federal de
     Itajub\'a, Itajub\'a MG, Brazil
     \and
     Instituto de Astronomia, Geof\'isica e Ci\^encias Atmosf\'ericas da Universidade de S\~ao Paulo,
Cidade Universit\'aria, CEP: 05508-900, S\~ao Paulo, SP, Brazil}

   \date{Received 6; accepted }
   \abstract{ The metallicity is a critical parameter that affects the correct
    determination fundamental characteristics stellar cluster  and
    has important implications in Galactic and Stellar evolution
    research.  Fewer than $10\%$ of the 2174 currently catalog open
    clusters have their metallicity determined in the literature.\\
    In this work we present a method for estimating the metallicity of
    open clusters via non-subjective isochrone fitting using the
    cross-entropy global optimization algorithm applied to UBV
    photometric data. The free parameters distance, reddening,
    age, and metallicity simultaneously determined by the fitting method.  The fitting procedure uses weights for the
    observational data based on the estimation of membership
    likelihood for each star, which considers the observational
    magnitude limit, the density profile of stars as a function of
    radius from the center of the cluster, and the density of stars in
    multi-dimensional magnitude space.\\
    We present results of [Fe/H] for nine well-studied open clusters
    based on 15 distinct UBV data sets.  The [Fe/H] values obtained in
    the ten cases for which spectroscopic determinations were available
    in the literature agree, indicating that our method
    provides a good alternative to determining [Fe/H] by
    using an objective isochrone fitting. Our results show that
    the typical precision is about 0.1~dex. }
        \titlerunning{Fitting Isochrones to Open Cluster}
\authorrunning{Oliveira et al.}
   \maketitle
 
 \section{Introduction}

The accurate determination of the fundamental parameters of open
clusters is essential in many fields of study in the Galactic and
stellar evolution context. Important questions that depend on
metallicity, which is usually measured by the [Fe/H] ratio, are the
determination of chemical abundance gradients (see \citet{lepine2011} and references therein), determination of the
rotational speed of the spiral pattern, and the co-rotation radius
\citep{Dias2005}, and in the stellar context the
empirical determination of the initial mass function, among many
other fields of study. 

Typically, the determination of distances, ages, and reddening of open
clusters via isochrone fitting requires either that the metallicity is
estimated, or that an priory value is adopted.  Therefore the 
  metallicity is a required parameter for the precise determinating
 the open cluster's fundamental characteristics. However, because of the
complexity of the observations required and the sometimes very
indirect methods needed to obtain this parameter, solar
metallicity is often assumed.

[Fe/H] can be estimated from spectroscopic data, from low-
to high-resolution, single or multi-object spectrographs as well as
from photometric data. Each technique has advantages and disadvantages and
limitations to the precision and accuracy that can be achieved. The
discussion of methods and techniques that allow the determination of
[Fe/H] is beyond the scope of this work, and we refer the reader the
reviews of \citet{gratton} and \citet{strobel}, among
others. For the estimates of [Fe/H] obtained via photometric data we
suggest the recent paper of \citet{paunzen_et_al} and
references therein.

In the last version of our open cluster catalog\footnote{Version 3.3
  of the new catalog of optically visible open clusters and candidates
  is available electronically at www.astro.iag.usp.br/\~{}wilton} (
\citet{dias2002} (DAML02)) we presented a compilation of [Fe/H] values
obtained from the literature for 202 open clusters. This compilation
is heterogeneous, since the metallicity determinations for a given
cluster were made from different data sets and techniques as well as
by different authors.  Of all clusters with metallicity estimates, in the
DAML02 catalog, we found that only 24\% of the objects have estimates
of their [Fe/H] ratio based on high-resolution spectra, 28\% are based
on low and medium-resolution spectra, and the rest are based on
photometric data. Of those, 28\% were estimated from isochrone
fitting. Values range from about -0.8~dex to +0.5~dex and the errors
range from 0.01~dex to 0.3~dex, depending on the method, number of
stars and techniques used. Note that in general there are no estimates
of the errors in [Fe/H] values obtained from isochrone fitting, and
for the six existing cases, the uncertainty varies from 0.15~dex to
0.50~dex.  Unfortunately, the subjectivity of the isochrone fitting
makes it difficult to estimate a reliable error of the [Fe/H] ratio.

It is a well known fact that estimates of [Fe/H] obtained by
high-resolution spectroscopy, which is the most reliable procedure,
are obtained from only a few stars for any given open cluster.  Due to
the dispendious nature of executing spectroscopy, previous selection
of the target stars is required, and thus the question of membership
determination becomes important.  Traditionally, the selection is made
based on the color-magnitude diagram (see for example the paper of
\citet{carrera}), choosing the red giant stars (usually the brightest
stars of the cluster), or considering a study of membership
probability based on proper motion and radial velocity data
(e.g. \citet{frinchaboy}). All these factors can seriously compromise
the determination of the [Fe/H] ratio if they are not made properly.

Given that only $9\%$ of all the 2174 catalog open clusters have
[Fe/H] estimated and that it is difficult to carry out detailed
spectroscopic study for a large number of stars in a large sample of
clusters, alternative and reliable methods are desirable.

In this work we focus on investigating the possibility of estimating
the metallicity of open clusters via isochrone fitting using the
cross-entropy global optimization algorithm (\citet{monteiro}, hereafter paper I), which allows simultaneous
determination of distance, reddening, age and metallicity. In the
second paper of this series (\citet{dias2012}, hereafter
paper II) we presented a nonparametric procedure to assign membership
likelihood based on photometric data of the stars, which in turn were
used as weights in the CE isochrone fitting.
To simplify the analysis in paper I and paper II  we kept the
metallicity constant at the value obtained from the literature which is
used by most previous studies. In this paper we use the metallicity as
a free parameter to be obtained from photometric UBV data and isochrone
fitting using the CE method.  In the next section we briefly review the CE
method and data used. In Sect. 3 we present the estimated metallicity
values obtained from the fitting method for each studied open
cluster. In the last section we conclude by emphasizing important
points, including potential applications and limitations of the work.
\section{Method and data}

In paper I we introduced a new technique to fit models to open cluster
photometric data using a weighted likelihood criterion to define the
goodness of fit and a global optimization algorithm known as
cross-entropy (CE) to find the best-fitting isochrone.

Very schematically, the CE procedure provides a simple adaptive way of
estimating the best-fit parameters. It involves an iterative
procedure that follows the steps outlined below:

\begin{itemize}
\item random generation of the initial sample of fit parameters,
  respecting predefined criteria;
\item selection of the best candidates based on calculated weighted
  likelihood values;
\item generation of a random fit parameter sample derived from a
  new distribution based on the previous step;
\item repeat until convergence or stopping criteria reached.

\end{itemize}

In paper II we introduced the nonparametric estimation of the
likelihood to obtain a better estimate of the probability whether a
given star is a member of the cluster.  The weighting scheme uses
observational data available in UBV filters for the open cluster and
calculates the membership likelihood for each star considering
observational magnitude limit, the density profile of stars as a
function of radius from the center of the cluster, and the density of
stars in multi-dimensional magnitude space. We refer to paper II for
more details.

As in paper I and paper II, the tabulated isochrones used were taken
from \citet{girardi} and \citet{marigo} and consisted of 400 files,
one for each isochrone, which are specified by two parameters, namely,
age and metallicity. To perform the isochrone fitting in this work we
included the parameters distance and reddening to define the parameter
space as follows:
\begin{enumerate}
\item {\bf Age}: from $log(age)=$6.60 to $log(age)=$10.15;  with steps of
      $log(age)=0.05$
  \item {\bf distance}: from 1 to 10000 parsecs;
  \item ${\bf E(B-V)}$: from 0.0 to 3.0;
  \item {\bf Metallicity}: from 0.0001 to 0.03~dex  with steps of
      $Z = 0.05 $~dex
\end{enumerate}

Applied our fitting procedure to the UBV data from the 
literature for the same open clusters analysed in paper I and paper
II. Apart from adopting metallicity as a free parameter, in this work
all procedures are identical to those in paper II.

To determine the parameter errors through Monte Carlo techniques we
performed the fit for each data set ten times, each time resampling
from the original data set, with replacement, to perform a bootstrap
procedure. In each bootstrap iteration new isochrone points from the
adopted initial mass function were also generated as described in paper
I, paper II, and in \citet{monteiro_dias}. The final
uncertainties in each parameter were then obtained by calculating the
standard deviation of the ten runs.

In previous papers of this series we have shown that the CE method was
robust, and the results obtained for the ten open clusters investigated
agree well with previous studies found in the literature, considering
UBV data (paper I), BVRI data \citep{monteiro_dias}
and also near-infrared (JH$K_{s}$) data obtained from the 2MASS catalog
(paper II). The method presents several advantages over visual fits,
especially since it removes most of the related subjectivity both in
the fit and in the weights of the stars in the color-magnitude diagram
(hereafter CMD), while also allowing us to determine the parameter
errors in a formal procedure.  The main limitation is that we do not
yet account for missing data. In other words, for a star to be
considered, it has to have been observed in all used filters. This
problem is being investigated and will be implemented for future
versions of the algorithm.

\section{Results and discussion}

As in papers, Table \ref{table0} presents the tuning parameters used
in the fitting procedure, where list the equatorial coordinates
($\alpha,\delta$) and the radius, which were obtained from the DAML02
catalog and $x_c$ and $y_c$ are the estimated center coordinates of
the cluster in the CCD, based on the determined 2D density
distribution of stars. The parameter $V_{cut}$ is the adopted cut-off
in V magnitude based on the completeness analysis in that band, and
$\mathbf{F}_{bin}$ is the binary fraction. The binary fraction was
changed in some cases to 50\% from the adopted 99\% where it clearly
improved the final fit. The paramenter $3\sigma_{phot}$ the
photometric error and $P_{cut}$ the adopted cut-off in weight
values. The WEBDA catalog\footnote{available at
  http://obswww.unige.ch/webda} \citep{Mermilliod1995} reference codes
are the same as those given in papers.

\begin{figure*} 
\centering
\includegraphics[scale=0.35]{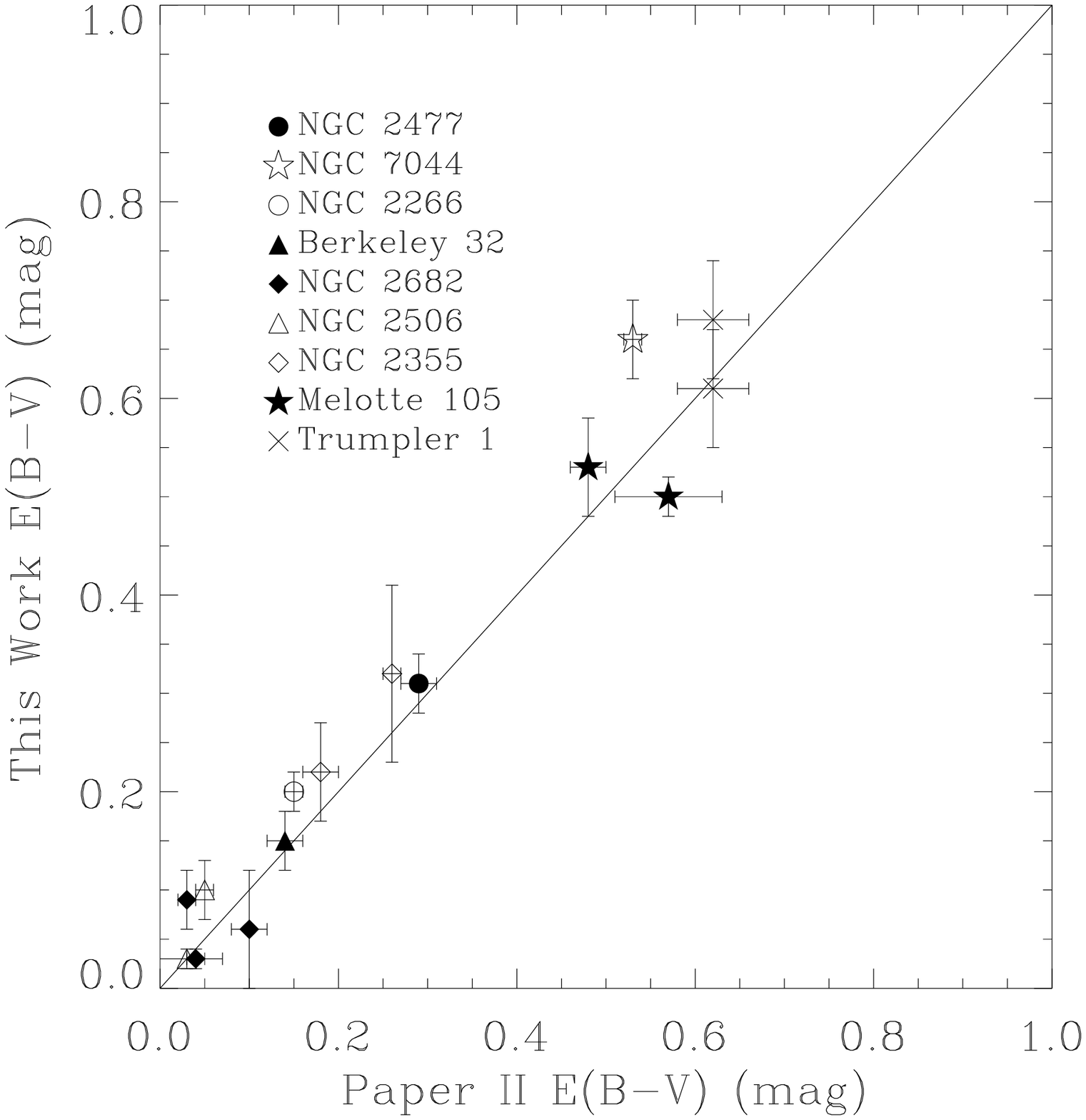}
\includegraphics[scale=0.35]{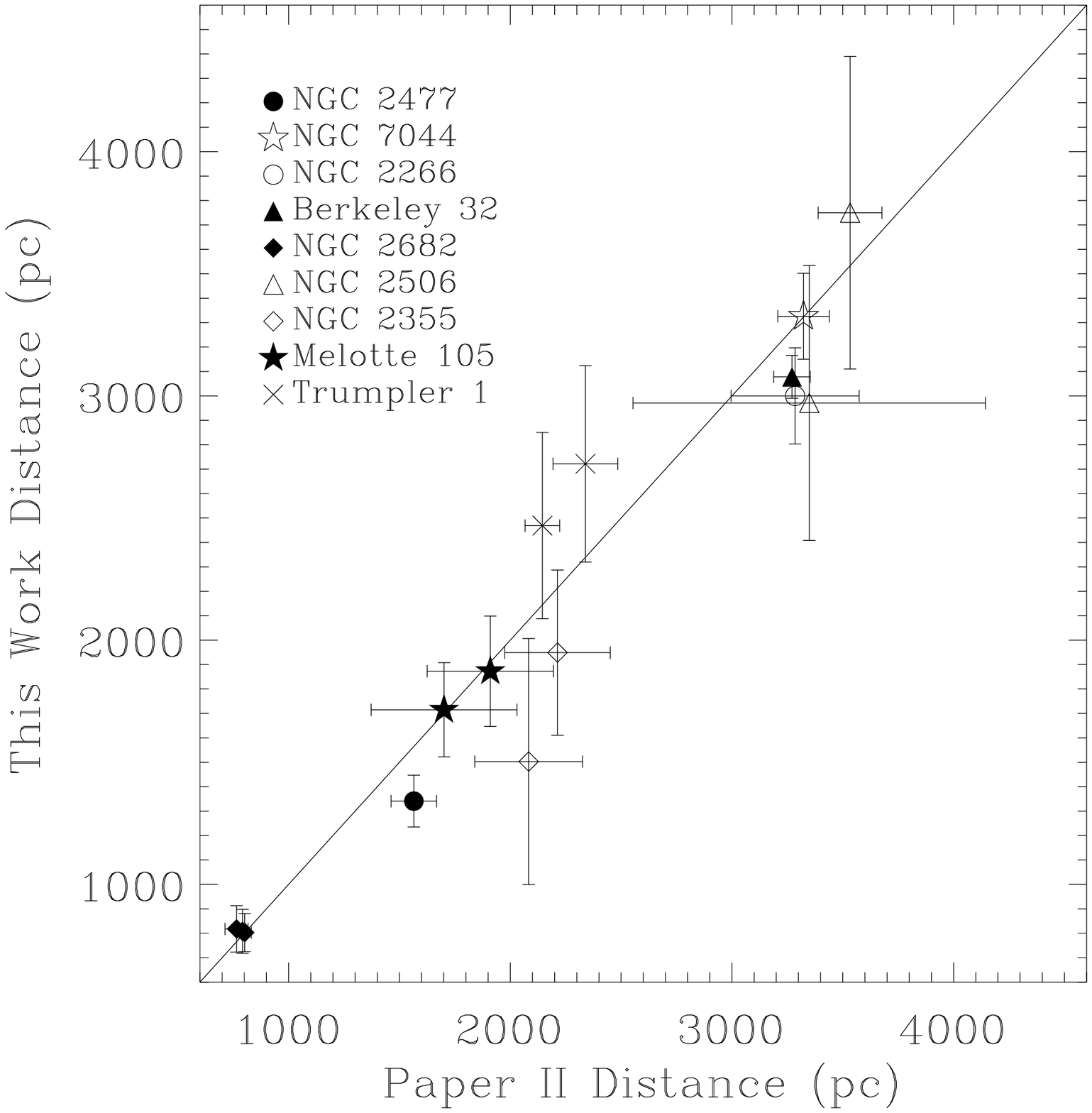}\\
\includegraphics[scale=0.35]{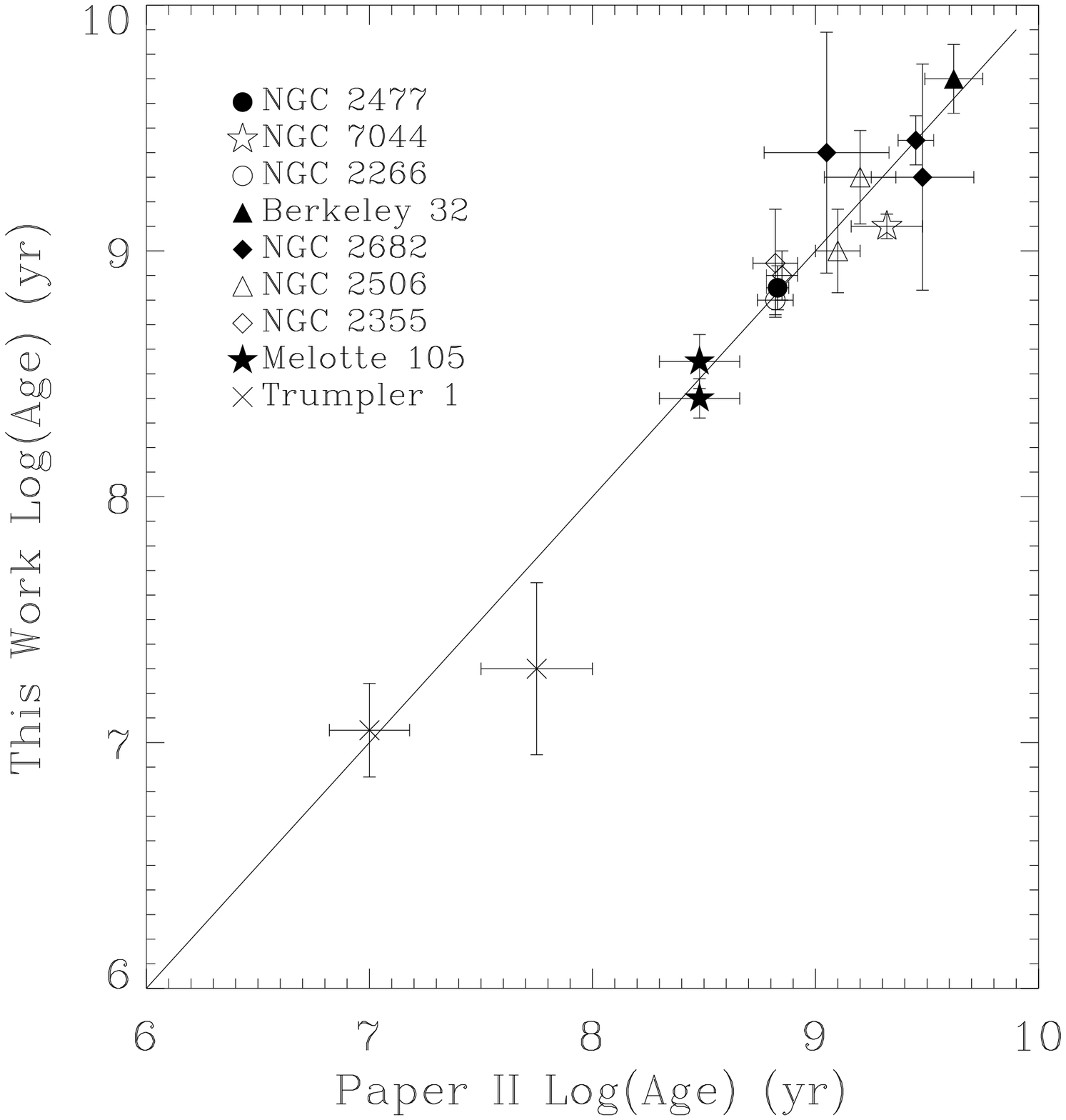}
\includegraphics[scale=0.35]{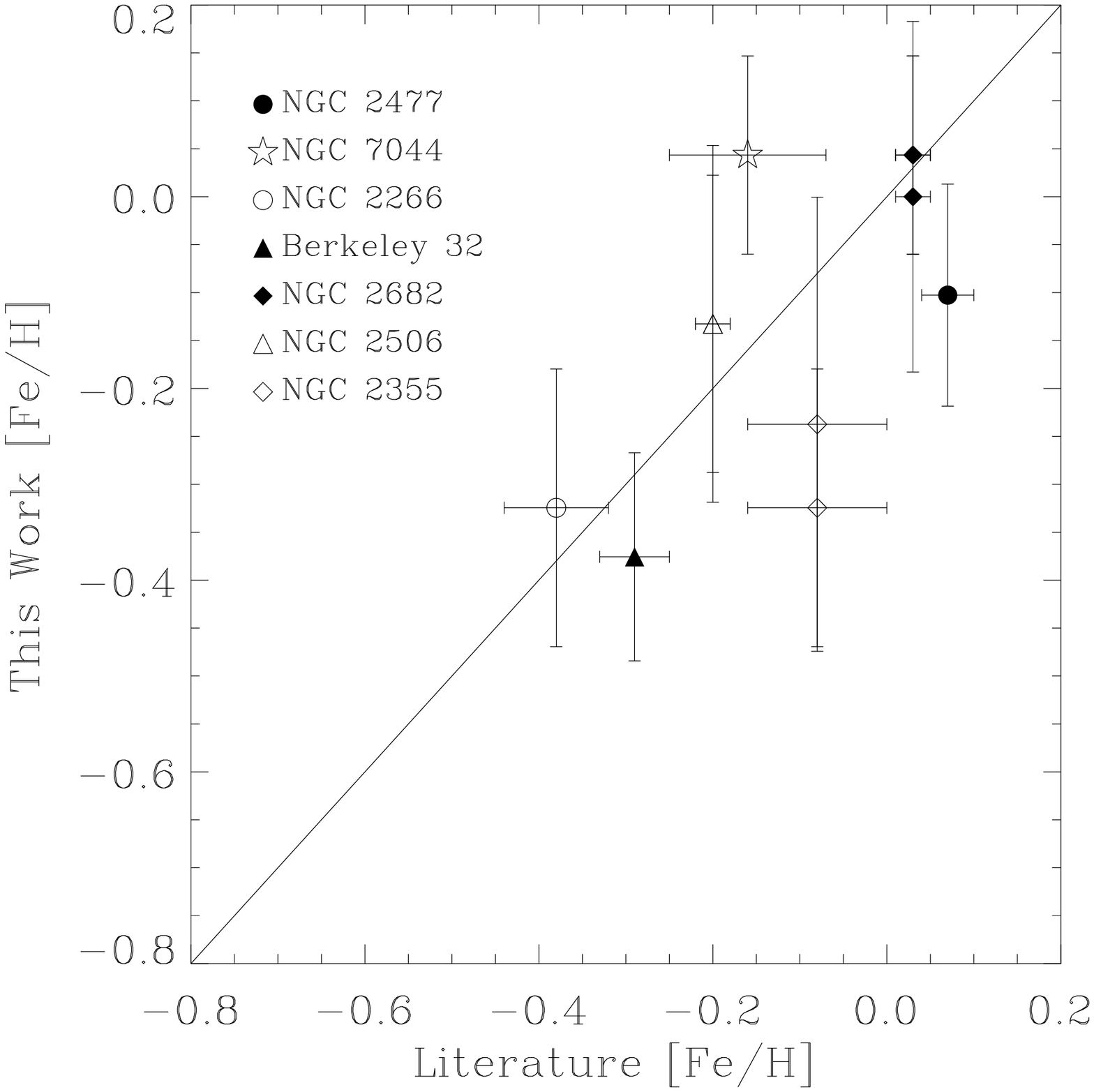}
\caption{Comparison of our fit results with those of paper II for 
    E(B-V), distance and age (upper left, right and lower left
    respectively). Comparison of final estimated metallicity values with those from the literature are given in the lower right plot. The error bars are those presented in
  Table \ref{table1} for the fits as explained in the text. The lines
  of 45$^\circ$ are the loci of equal values. The [Fe/H] values
  obtained are compared with those from the DAML02 catalog. The filled
  symbols indicate [Fe/H] from high-resolution spectra.}
\label{comp1}
\end{figure*}

We present the comparison of the results obtained in this work with
those obtained in paper II in the first three plots of Fig.~\ref{comp1}
for the nine open clusters. The last plot shows the metallicity
values we determined in this work compared with literature values from the
DAML02 catalog, which were obtained from spectroscopy.

The average and standard deviation of the differences
of our results to those of paper II are
\begin{description}
\item{ {\bf E(B-V)}}= $0.03\pm0.05$~$mag$;
\item{ {\bf Distance}}=       $-63\pm      262$~$pc$;
\item{ {\bf Log(age)}}=        $-0.01 \pm       0.18$~$yr$.
\end{description}

One can see from the comparison of the results that distance, E(B-V) and
age values obtained in paper II were recovered in this work, within
the uncertainties of the method.

The fit results obtained by the method applied to the UBV data for
each cluster can be seen in Figs.~\ref{fig3} through \ref{fig17} in
Appendix A. The figures present the CMDs with the original data
followed by the same plots, with the symbol sizes reflecting the
weights used as obtained from the procedure introduced in paper
II. The fitted isochrone and zero age main sequence are also
plotted. In Tables~\ref{table1} and~\ref{table2} we present the final
fit values for each cluster studied with the metallicity as a free
parameter in the fitting procedure. To facilitate the comparison, the
parameter values obtained in paper II are also provided, including the
metallicity adopted from the literature. The final value of this
parameter was transformed to [Fe/H], adopting the same approximation
as considered in the Padova database of stellar evolutionary tracks
and isochrones: $[Fe/H] = logZ/Z\odot$ with $Z\odot = 0.019$.  The
errors were obtained by the usual propagation formula.

In Table~\ref{table2} our results for [Fe/H] are compared with those
from the literature where we also provide information on the method
used in the given reference to determine it.  In comparing our results
with those in the literature that used spectroscopy to obtain [Fe/H], we
find that the average difference is 0.08~dex with a standard deviation
of 0.07~dex, with no significant difference between regular or high-resolution spectroscopy (HRS).

The [Fe/H] values we determine agree with those obtained
from the literature considering the previous comparison. The mean of
the differences shows that there is no  significant systematic
trend, and the low value of the mean square difference indicates
that both sets of measurements agree.  Considering that the
values obtained by HRS are the most reliable, the agreement of our
values with those in the literature indicates that our method provides
adequate results for [Fe/H] by isochrone fitting. It would be
interesting to fit a large number of clusters to confirm this as well
as to allow investigation of possible biases.

Below we comment on some individual
open cluster results.

\subsection{\object{NGC~2477}}

The metallicity determined from our fitting procedure for the cluster
NGC~2477 is just outside of the $1\sigma$ agreement region, which can
be seen in Fig. \ref{comp1}, considering [Fe/H] values from DAML02
catalog. In DAML02 the value from \citet{Bragaglia2008} was adopted
estimated on the basis of six stars from HRS. Comparing their results
with those from \citet{friel_2002}, the authors consider, the
possibility that their metallicities could be generally bigher by
about 0.2~dex.  The values in Table~\ref{table2} indicate the
possibility of lower metallicity for \object{NGC~2477}.  Considering
the original photometric data \citep{Kassis1997} (REF 152), our
results agree with those obtained by the authors, who used
$[Fe/H]=-0.05\pm 0.11 $~dex from moderate resolution spectroscopy of
seven cluster giants determined by \citet{friel_janes}.  Another
interesting point is that our values agree with those of Jeffery et
al. (2011). In that study the authors employed a new Bayesian
statistical technique that performs an objective, simultaneous model
fit of the cluster and stellar parameters with the photometry. The
authors used BVI photometric data, obtaining E(B-V) = 0.198, distance
of 1411 pc and logt of 9.04, and $[Fe/H]= -0.34 \pm 0.07 $~dex.  The
[Fe/H] value estimated by our fit agrees within the uncertainties with
the values published previously by \citet{friel_janes} and
\citet{jeffery}.  Possibly a more complete UBVRI data set togheter
with the CE algorithm could confirm the value of [Fe/H] for the
object.


\subsection{\object{NGC~2355}}

\begin{figure}[h] 
\centering
\includegraphics[scale=0.4]{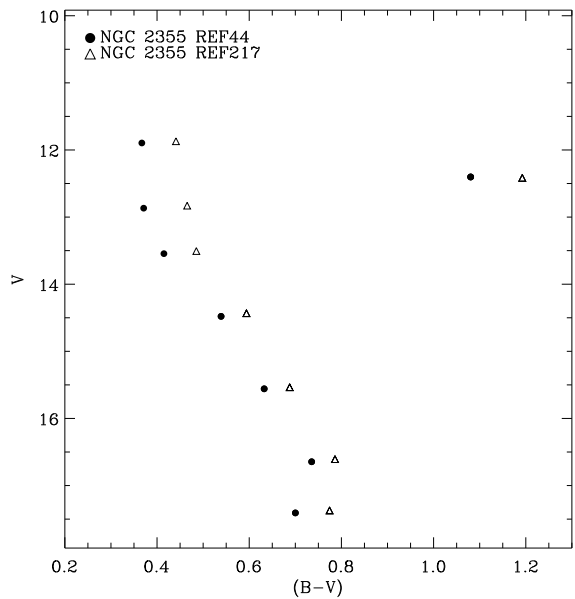}
\caption{Comparison of average color-index values for the two data
  sets used for the open cluster \object{NGC~2355}. The differences in
  data sets explain the differences in the parameters obtained with
  the cross-entropy method, specially distance and metallicity.}
\label{syserr}
\end{figure}

The open cluster \object{NGC~2355} also shows significant differences
in the estimated metallicity with the data set from \citet{Ann1999}
(REF. 217) when compared with literature values obtained from
spectroscopy. Despite the obtained errors in [Fe/H], the discrepancy in
this case seems to be due to a systematic difference in the photometry
of the two sets. The data from \citet{Ann1999} are systematically
redder than the one from \citet{Kaluzny_c} (REF. 44). To show
this difference, we took the two data sets and calculated color-index
averages for specific V-magnitude bins.  The stars used in each
  bin from each data set are the same, to avoid biases and selection
  effects. We then plotted the values in the CMD. For the locus where
the giants are, we defined a box such that $10<V<14$ and $(B-V)>0.8$ to
obtain the color-index average. For the rest of the data we used a V-magnitude bin of 1. The result is shown in Fig.\ref{syserr}.  This
example is useful to illustrate that the quality of the [Fe/H]
estimate obtained by fitting is directly linked to the quality of the data, even considering the improved statistical fitting
procedures.

\subsection{\object{NGC~7044}}
 
\begin{figure}[h] 
\centering
\includegraphics[scale=0.4]{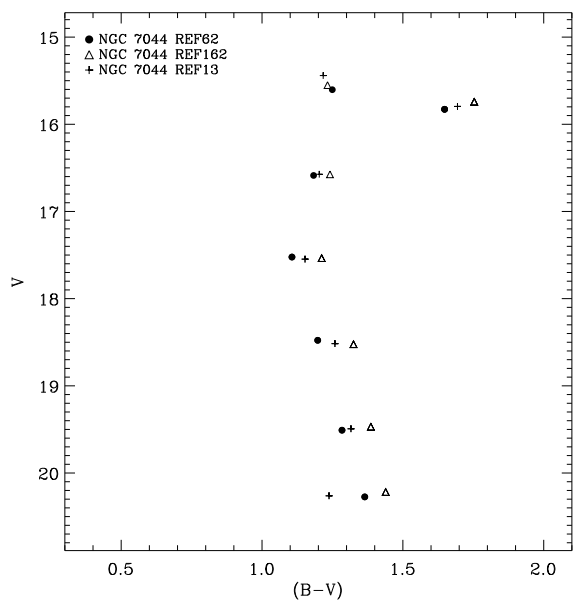}
\caption{Comparison of mean color-index values for the data set used in the fit of the open cluster \object{NGC~7044} and the data sets of \cite{SG98} (REF~162) and \cite{K89} (REF~13). The differences in the data sets are a possible cause of the discrepancy between the value obtained for the metallicity by our fit and the one of Warren \& Cole (\cite{warren_cole}), which was obtained with spectroscopy. }
\label{n7044comp}
\end{figure}

Our result for the open cluster \object{NGC~7044} also shows a
significant difference from the result of  \cite{warren_cole} obtained
using spectroscopy.  Unlike the case for \object{NGC~2355}
mentioned before, here we did not have a second data set with UBV
photometry to compare distinct fit results. However, we were able
to compare the observational data by using the B and V values
obtained by \cite{K89} and \cite{SG98}. To perform the comparison,
we followed the same strategy as before where we calculated color-index averages for specific V-magnitude bins in the CMD  considering the same stars. For the
giant locus in the CMD of \object{NGC~7044} we defined a box
such that $14<V<17$ and $(B-V)>1.5$ to obtain the color-index
average. For the rest of the data we used a V-magnitude bin of
1. The result of the comparison is shown in Fig. \ref{n7044comp},
where it is clear that there is a considerable difference in the
photometry.  Even though we were not able to perform fits to the
data from \cite{K89} and \cite{SG98} since they only observed B
and V filters, it is likely that the photometric differences
shown are an important factor in accounting for the discrepancy
in metallicity that we found.

\begin{table*}
  \caption{Cross-entropy fit parameters. The first five columns (after the cluster identification) give the central coordinates and radius from the DAML02 catalog, followed by the X,Y central position used considering UBV data from the literature. The parameter $V_{cut}$  is the adopted cut-off in magnitude in V, $\mathbf{F}_{bin}$ is the number of stars considered as binary, $3\sigma_{phot}$ the photometric error, and $P_{cut}$ the adopted normalized likelihood cut-off. The reference codes given in last column are the same as used by WEBDA and were also used in paper I. }
\tiny
\label{table0}
\centering
\begin{tabular}{l c c c c c c c c c c}
\hline\hline
Cluster &  {$\alpha$}  & {$\delta$} & RADIUS & $x_c$ & $y_c$ & $V_{cut}$& $\mathbf{F}_{bin}$& $3\sigma_{phot}$&$P_{cut}$  & REF\\
        & J2000.0      &  J2000.0  &  arcmin  &  pix & pix & (mag) & (\%) &(\%)&(\%) & \\
\hline
\object{NGC~2477}    &07 52 10     & -38 31 48           &15 	 & 	-40.97	 & 	-160.82	 &	18.25	 &50   	&1.0 &	5.0 	  &152	\\
\object{NGC~7044}    & 21 13 09    & +42 29 42     	&06  	 & 	13.27	 &	31.35	 &		 &99	&1.5 &	10	  &62	\\
\object{NGC~2266}    & 06 43 19    &  +26 58 12    	&05  	 &	0.05	 &	7.94	 &		 &99	&1.0 &	5.0	  &41	\\
\object{Berkeley 32} & 06 58 06    &   +06 26 00   	&06  	 &	734.22	 &	662.30	 &		 &99	&1.0 &		  &40	\\
\object{NGC~2682}    & 08 51 18    &  +11 48 00    	&25  	 &	-0.01	 &	-0.43	 &		 &99	&1.5 &		  &335	\\
            &             &               	&  	 &	-0.22	 &	-0.83	 &		 &50	&1.5 &		  &31	\\
            &             &               	&  	 &	0.00	 &	-1.64	 &		 &50	&1.5 &	5.0	  &54	\\
\object{NGC~2506}    & 08 00 01    & -10 46 12     	&12  	 &	-9.58	 &	-16.10	 &		 &99	&1.0 &	30	  &284	\\
            &             &               	&  	 &	-10.85	 &	8.30	 &	17.75	 &99	&1.5 &	5.0	  &163	\\
\object{NGC~2355}    & 07 16 59    &  +13 45 00    	&07  	 &	33.30	 &	12.38	 &		 &99	&1.5 &		  &217	\\
            &             &               	&  	 &	17.97    &	3.40	 &		 &99	&1.5 &	5.0	  &44	\\
\object{Melotte~105} & 11 19 42    &  -63 29 00    	&05  	 &	-39.31	 &	-35.96	 &		 &99	&1.5 &	5.0	  &289	\\
            &             &               	&  	 &	11.18	 &	-34.34	 &	16.75	 &99	&1.5 &		  &32	\\
\object{Trumpler~1}  & 01 35 42    &  +61 17 00    	&03  	 &	1.54	 &	-2.04	 &	17.75 	 &99	&1.5 &	5.0	  &320	\\
            &             &               	&  	 &	-19.05	 &	0.68	 &	17.75	 &99	&1.0 &		  &86            	\\
\hline
\end{tabular}
\begin{flushleft} 
\tiny
References: \\
152 =  \cite{Kassis1997} \\
62  =  \cite{Aparicio1993}\\ 
41  =   \cite{kaluzny_a}  \\
40  =  \cite{kaluzny_b} \\ 
335 =  \cite{henden2003}  \\
31  =  \cite{Gilliland1991} \\
54  =  \cite{Montgomery1993} (adopted logt = 9.6)\\
284 =  \cite{Kim2001} (adopted mean values: see Table 5 of the paper) \\    
163 =  \cite{Marconi1997}\\  
217 =  \cite{Ann1999}  \\
44  =  \cite{Kaluzny_c}\\   
289 =  \cite{Sagar2001}\\
32  =  \cite{Kjeldsen1991}  \\
320 =  \cite{Yadav2002}\\ 
86  =  \cite{Phelps1994}\\
\end{flushleft}
\end{table*}

\begin{table*}
  \caption{Parameters obtained for the investigated clusters with the cross-entropy method considering the metallicity as a free parameter. In the first three columns (after the cluster identification) we reproduce the results published in paper II to facilitate comparison. In the following three columns the results for $E(B-V)$ the extinction, distance to the cluster and $\log(Age)$ the logarithm of the age (in years) obtained in this work are presented. The reference codes given in last column are the same as given in Table \ref{table0}. See the text for error estimate details.}
\tiny
\label{table1}
\centering
\begin{tabular}{l c c c c c c c c c}
\hline\hline
&\multicolumn{3}{c}{paper II}& \multicolumn{3}{c}{This work}&\\
Cluster & $E(B-V)$ & Distance & Log(Age) & $E(B-V)$ & Distance & Log(Age)& REF\\
        & (mag)&(pc)&(yr)&(mag)&(pc)&(yr)&\\
\hline
 \object{NGC~2477}    &0.29 $\pm$	0.02&	1565 $\pm$	103&	8.83 $\pm$	0.05& 0.31 $\pm$ 	 0.03&	 1341 $\pm$	106&	 8.85 $\pm$	0.09 	&152		\\
 \object{NGC~7044}    &0.53 $\pm$	0.01&	3323 $\pm$	116&	9.32 $\pm$	0.16& 0.66 $\pm$	 0.04&	 3326 $\pm$	176&	 9.10 $\pm$	0.05	&62	 	\\
 \object{NGC~2266}    &0.15 $\pm$	0.01&	3285 $\pm$	289&	8.82 $\pm$	0.08& 0.20 $\pm$	 0.02&	 3000 $\pm$	197&	 8.80 $\pm$	0.06	&41	 	\\
 \object{Berkeley 32} &0.14 $\pm$	0.02&	3271 $\pm$	83&	9.62 $\pm$	0.13& 0.15 $\pm$	 0.03&	 3078 $\pm$	88&	 9.70 $\pm$	0.14	&40	 	\\
 \object{NGC~2682}    &0.03 $\pm$	0.01&	765  $\pm$	52&	9.48 $\pm$	0.23& 0.09 $\pm$	 0.03&	 818  $\pm$	95&	 9.30 $\pm$	0.46 	&335		\\
             &0.10 $\pm$	0.02&	802  $\pm$	30&	9.05 $\pm$	0.28& 0.06 $\pm$	 0.06&	 803  $\pm$	78&	 9.40 $\pm$	0.49	&31	 	\\
             &0.04 $\pm$	0.01&	792  $\pm$	20&	9.45 $\pm$	0.08& 0.03 $\pm$	 0.01&	 808  $\pm$	90 &	 9.45 $\pm$	0.10	&54		\\
 \object{NGC~2506}    &0.03 $\pm$	0.04&	3349 $\pm$	795&	9.20 $\pm$	0.16& 0.03 $\pm$	 0.01&	 2970 $\pm$	563&	 9.30 $\pm$	0.19	&284	 	\\
             &0.05 $\pm$	0.01&	3533 $\pm$	144&	9.10 $\pm$	0.10& 0.10 $\pm$	 0.03&	 3750 $\pm$	640&	 9.00 $\pm$	0.17	&163	 	\\
 \object{NGC~2355}    &0.26 $\pm$	0.01&	2083 $\pm$	243&	8.82 $\pm$	0.10& 0.32 $\pm$	 0.09&	 1503 $\pm$	504&	 8.95 $\pm$	0.22	&217		\\
             &0.18 $\pm$	0.02&	2213 $\pm$	238&	8.85 $\pm$	0.07& 0.22 $\pm$	 0.05&	 1949 $\pm$	338&	 8.90 $\pm$	0.10	&44		\\
 \object{Melotte~105} &0.48 $\pm$	0.02&	1701 $\pm$	329&	8.48 $\pm$	0.18& 0.53 $\pm$	 0.05&	 1715 $\pm$	193&	 8.55 $\pm$	0.11	&289	 	\\
             &0.57 $\pm$	0.06&	1910 $\pm$	285&	8.48 $\pm$	0.18& 0.50 $\pm$	 0.02&	 1873 $\pm$	226&	 8.40 $\pm$	0.08	&32		\\
 \object{Trumpler~1}  &0.62 $\pm$	0.04&	2145 $\pm$	78& 	7.75 $\pm$	0.25& 0.68 $\pm$	 0.06&	 2469 $\pm$	381&	 7.30 $\pm$	0.35	&320	 	\\
             &0.62 $\pm$	0.04&	2339 $\pm$	146&	7.00 $\pm$	0.18& 0.61 $\pm$	 0.06&	 2722 $\pm$	402&	 7.05 $\pm$	0.19	&86		\\
\hline
\end{tabular}
\end{table*}


\begin{table*}
  \caption{Parameter metallicity obtained for the clusters investigated with the cross-entropy method. In the first column we list the cluster identification, the second column shows the metallicities obtained from the literature for the investigated clusters, and the third column presents the literature codes. In the column TEC we give the technique of data acquisition used for [Fe/H] determination, named SPEC for spectroscopy and PHOT for photometry. The last two columns present the metallicity values obtained in this work and the reference code for the data used from the WEBDA catalog. See the text for error estimates details.}
\tiny
\label{table2}
\centering
\begin{tabular}{l c c c  c c}
\hline\hline
&\multicolumn{3}{c}{}& \multicolumn{2}{c}{This work}\\
Cluster & [Fe/H] & literature code & TEC code & [Fe/H] & REF \\ 
\hline
 \object{NGC~2477}    & 0.07  $\pm$ 0.03 & R01 & SPEC  	& 	-0.10  $\pm$ 0.12    &  152	\\
             & -0.03$\pm$ 0.07  &R03  & PHOT    &	  &	 \\
             & -0.008   &R07  & PHOT   &	  &	 \\
             & -0.13$\pm$ 0.18  &R08  & PHOT    &	  &	 \\
             & -0.34$\pm$ 0.07  &R09  & PHOT   &	  &	 \\  
             & -0.05$\pm$ 0.11 &R10  & SPEC    &	  &	 \\ 
             & 0.04$\pm$ 0.01  &R11  & SPEC   &	  &	 \\ 
             & 0.019$\pm$ 0.115  &R12  &SPEC    &	  &	 \\ 
             & 0.05  &R13  & PHOT    &	  &	 \\  
              & 0.07$\pm$ 0.03  &R28  & SPEC    &	  &	 \\      
 \object{NGC~7044}    &  -0.16 $\pm$ 0.09                &  R14   &    SPEC 	& 	 0.04 $\pm$ 0.10   & 62 	\\
            &  0.0  $\pm$ 0.2   &  R27   &    PHOT 	& 	   &  	\\
            &  0.01                &  R13   &    PHOT 	& 	   &  	\\
             & 0.01$\pm$ 0.10  &R03  & PHOT   &	  &	 \\ 
 \object{NGC~2266}    & -0.26 $\pm$ 0.02          & R02 & PHOT 	& 	-0.32  $\pm$ 0.14   & 41 	\\
             &  -0.26 $\pm$ 0.2  &R03  & PHOT  &	  &	 \\ 
             &  -0.38 $\pm$ 0.06  &R15  & SPEC  &	  &	 \\ 
 \object{Berkeley 32} & -0.29 $\pm$ 0.04 & R01 & SPEC 	& 	-0.38  $\pm$ 0.11& 40 	\\
             &  -0.42 $\pm$ 0.09  &R03  & PHOT  &	  &	 \\ 
             &  -0.37 $\pm$ 0.05  &R16  & PHOT &	  &	 \\ 
             &  -0.3 $\pm$ 0.02  &R17  & SPEC  &	  &	 \\ 
             &  -0.37 $\pm$ 0.04  &R18  & PHOT  &	  &	 \\ 
             &  -0.55   &R13  & PHOT  &	  &	 \\ 
 \object{NGC~2682}    &  0.03 $\pm$ 0.02   & R04 & SPEC 	&	  0.0  $\pm$ 0.18&	335	\\
             &  -0.029            & R07 & PHOT 	& 	 0.04  $\pm$ 0.10& 31 	\\
             &  -0.04$\pm$ 0.03   & R03 &  PHOT 	& 	0.04  $\pm$ 0.10 &  54    \\
             &  -0.05 $\pm$ 0.03  &R19  & PHOT  &	  &	 \\ 
             &  -0.06 $\pm$ 0.07  &R20  & PHOT   &	  &	 \\ 
             &  -0.05 $\pm$ 0.04  &R18  & PHOT  &	  &	 \\ 
             &  -0.01 $\pm$ 0.11  &R21  & PHOT   &	  &	 \\ 
             &  0.000 $\pm$ 0.092   &R12 & PHOT   &	  &	 \\ 
             &  -0.11             &R13  & PHOT  &	  &	 \\ 
             &  -0.09 $\pm$ 0.07  &R10  & SPEC  &	  &	 \\ 
             &  -0.01 $\pm$ 0.05  &R06  & SPEC   &	  &	 \\             
 \object{NGC~2506}    & -0.2  $\pm$ 0.02   &R05  & SPEC  	& 	-0.13  $\pm$ 0.16& 	284\\
             &  -0.51 $\pm$ 0.1   &R03  &  PHOT 	& 	-0.13  $\pm$ 0.16& 163	\\
             &  -0.19 $\pm$ 0.06  &R22  & SPEC   &	  &	 \\
             &  -0.24 $\pm$ 0.05  &R23  & SPEC   &	  &	 \\
             &  -0.44 $\pm$ 0.06  &R24  & SPEC   &	  &	 \\
             &  -0.58 $\pm$ 0.14  &R08  & PHOT   &	  &	 \\
             &  -0.52 $\pm$ 0.07  &R10  & SPEC  &	  &	 \\
             &  -0.57             &R25  & PHOT   &	  &	 \\
             &  -0.48 $\pm$ 0.08  &R21  & PHOT   &	  &	 \\
             &  -0.58             &R13  & PHOT  &	  &	 \\
             &  -0.368 $\pm$0.108 &R12  & PHOT   &	  &	 \\
 \object{NGC~2355}    & -0.08 $\pm$ 0.08   & R06 & SPEC  	&	  -0.32  $\pm$ 0.14&	217\\
             & 0.02$\pm$0.2       & R03 & PHOT	&	  -0.23  $\pm$ 0.24&	44 \\
             & 0.13               &R26  & PHOT   &	  &	 \\
 \object{Melotte~105} &  0.0               & R13 & PHOT 	& 	-0.05  $\pm$ 0.15& 289	\\
             &  0.0  $\pm$ 0.1    & R03 &  PHOT 	&	  0.00   $\pm$ 0.21&	32 \\
 \object{Trumpler~1}  &  -0.71             &  R13&  PHOT 	& 	0.10  $\pm$ 0.13& 320	\\
             &  -0.71  $\pm$ 0.1  &  R03 &   PHOT   	&	  0.15  $\pm$ 0.10&	86 \\
\hline
\end{tabular}
\tablefoot{The TEC codes are
R01, R04, R17, R22, and R28  used the data from high-resolution spectroscopy; 
R03 used unweighted averaged [Fe/H] values from the literature; 
R05, R06, R10, R11, R12, R14, R15, R23 and R24 used the data from high-, moderate- or low-resolution spectroscopy; 
R07, R09, R16, R18, R19, R20, R25, R26 and R27 used values from color-magnitude isochrone fits; 
R02, R08 and R16 used values from Washington photometry; 
R13 used UV excesses techniques ($\delta(U-B)_{0.6}$); 
R12, R19 and R20 used values from DDO photometry, and 
R21 used recalibrated values from Piatti et al. (\cite{piatti})  \\
Literature code: \\
R01 = \citet{sestito}; R02 = \citet{kaluzny_a}; R03 = \citet{paunzen_et_al}; R04 = \citet{randich}; R05 = \citet{carretta}; R06~=~\citet{jacobson}; R07 = \citet{cameron}; R08 = \citet{geisler}; R09 = \citet{jeffery}; R10 = \citet{friel_janes}; R11 = \citet{smith_hesser}; R12 = \citet{twarog}; R13 = \citet{tadross}; R14 = \citet{warren_cole}; R15 = \citet{carrera}; R16 = \citet{kaluzny_b};  R17 = \citet{carrera_2011}; R18 = \citet{noriega}; R19 = \citet{janes}; R20 = \citet{nissen}; R21 = \citet{piatti}; R22 = \citet{reddy}; R23 = \citet{Mikolaitis}; R24 = \citet{friel_2002};  R25 = \citet{mcclure}; R26 = \citet{kaluzny_c}; R27 = \citet{Aparicio1993}; R28 = \citet{Bragaglia2008}  } 
\end{table*}

\section{Conclusions}
\label{con}

The observational complexity and difficulty in carrying out detailed
high-resolution spectroscopy of a large number of stars for a large
number of clusters raises the question of alternative methods for
estimating their metallicity reliably . The observational complexities
account for the very small number of clusters, fewer than 10\% in
DAML02 catalog, for which good-quality metallicities exist. As
commented by \citet{paunzen}, the metallicity parameter is set as
solar or ignored in most papers that perform some sort of isochrone
fitting in CMDs when this parameter is not available from other
sources, possibly introducing an unknown bias in the distance, age and
reddening estimated.

Our method, based on the Cross-Entropy optimization algorithm, using
UBV photometric data weighted with a membership-likelihood estimation,
allows for the simultaneous determination of the parameters distance,
reddening, age and, as shown in this work, metallicity. The main
advantage, as discussed in detail in the previous works of the series,
is the automation and removal of subjectivity in the fitting
process. All fitting parameters are clearly defined and the fitting
quality is quantifiable through the likelihood, which can be
objectively compared with any proposed alternative.

In this work we present [Fe/H] estimates obtained with the CE method
for nine well-studied open clusters based on 15 distinct UBV data
sets. The comparison with [Fe/H] obtained by spectroscopy indicates
that our results are adequate for obtaining low-telescope-cost
metallicity estimates since most values are consistent withing the
$1\sigma$ uncertainty. Our bootstrap-estimated errors for [Fe/H]
values also show the good agreement between our results and
literature [Fe/H] values obtained from photometric as well as
spectroscopic data.

It is important to point out that the accuracy and precision as well
as the overall quality of the photometric data are decisive for the
quality of the final estimated metallicity. 

Generally, visual isochrone fits can provide estimates of the
metallicity, but typically do not provide estimates of the error (in
some works the error estimates are given as in \citet{vazquez}, although obtained by visual fit). Our method
provides a robust [Fe/H] with consistent error estimates.  Our results show a typical internal precision of about 0.1~dex.

Finally, we emphasize the conclusion of paper I, but now with the
extra parameter metallicity included, that our method is reliable and
robust in determining the distance, age, reddening, and metallicity by
isochrone fits. It is a powerful tool that in the near future can be
used with already existing data and especially in upcoming modern
large surveys, such as GAIA, VISTA, and Pan-STARRS to produce
homogeneous large samples of determined fundamental parameters of open
clusters.
\begin{acknowledgements}

  We acknowledge support from the Brazilian Institutions CNPq and
  INCT-A. H. Monteiro acknowledges the CNPq (Grant 470135/2010-7) and
  FAPEMIG (Grant APQ-02030-10 and CEX-PPM-00235-12). T. Caetano thanks
  CNPq for financial support. We also made use of the WEBDA open
  cluster database.

\end{acknowledgements}   
\bibliographystyle{aa} 
\bibliography{ce_opt_refs.bib} 
\begin{appendix}
\label{appe}
\section{Fit results}
  \begin{figure*}[!h]
\centering
\includegraphics[width=18cm]{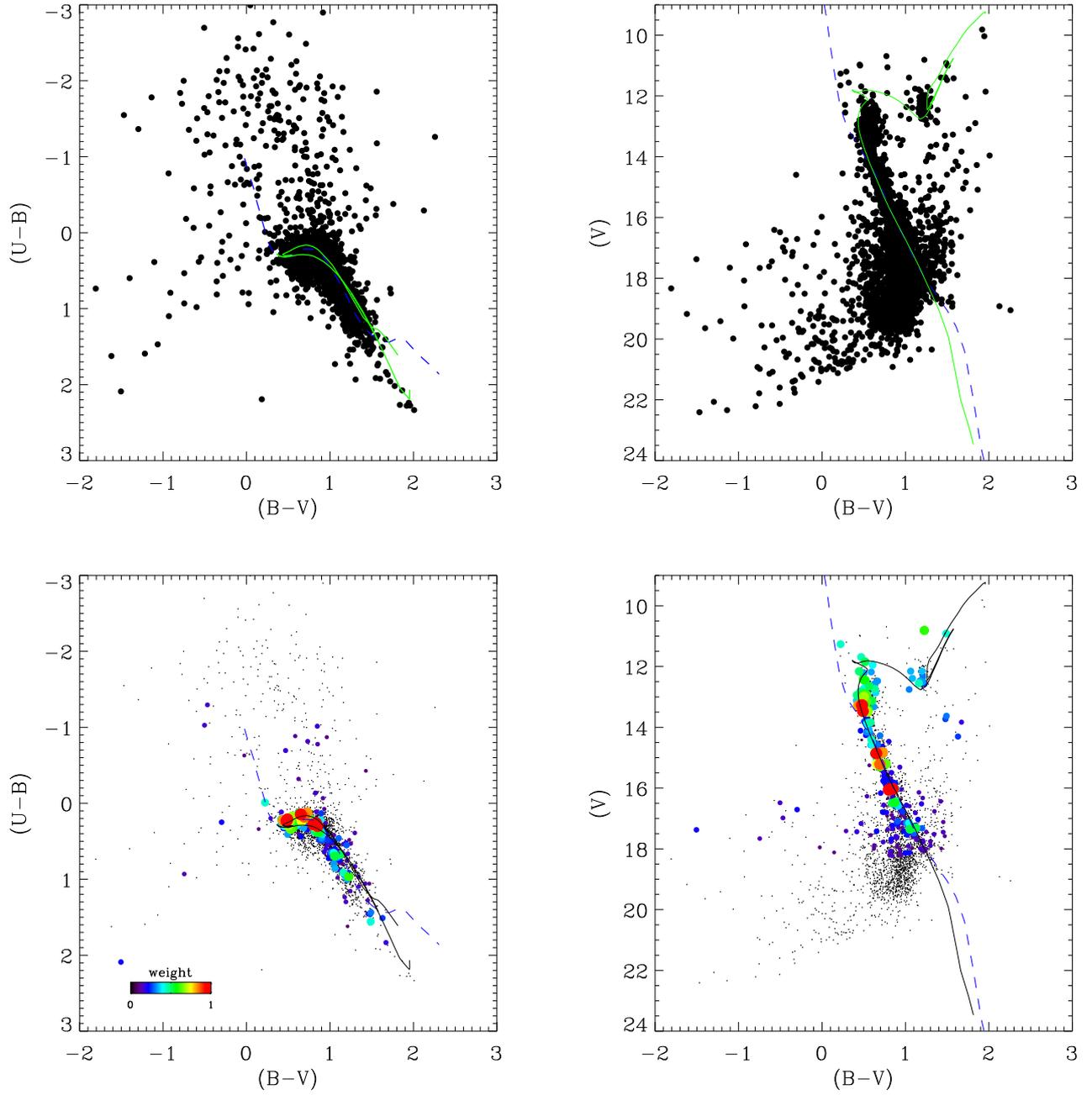}
\caption{Results for the open cluster NGC~2477. Upper graphs show 
   available UBV data and lower ones the weighted data (with symbol size and color ranging proportional to membership likelihood), the fitted isochrone (solid line) and the ZAMS (dashed line).}
\label{fig3}
\end{figure*}

\begin{figure*}
\centering
\includegraphics[width=18cm]{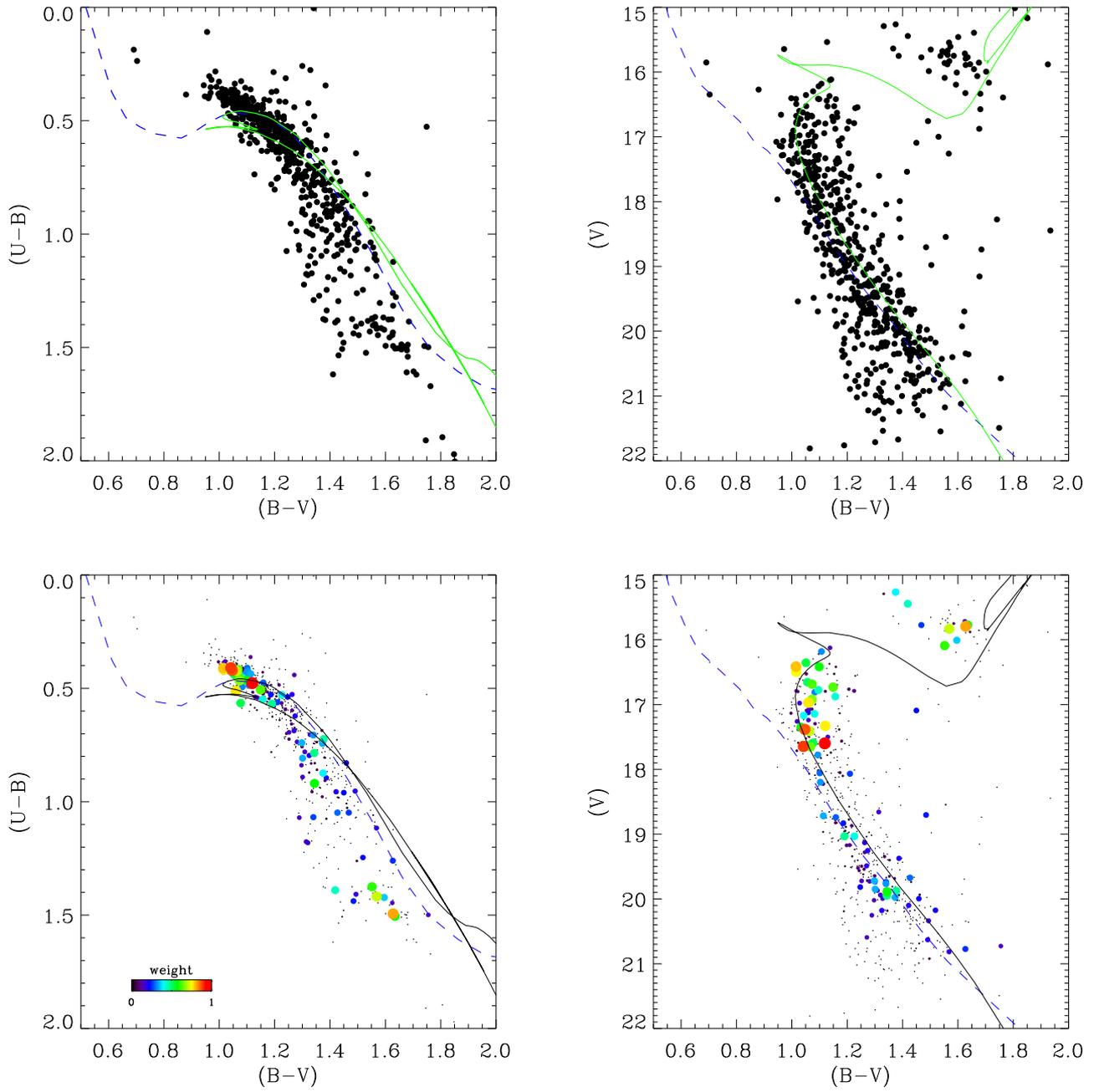}
\caption{Same as Fig.~\ref{fig3} for NGC~7044. }
\label{fig4}
\end{figure*}
 
\begin{figure*}
\centering
\includegraphics[width=18cm]{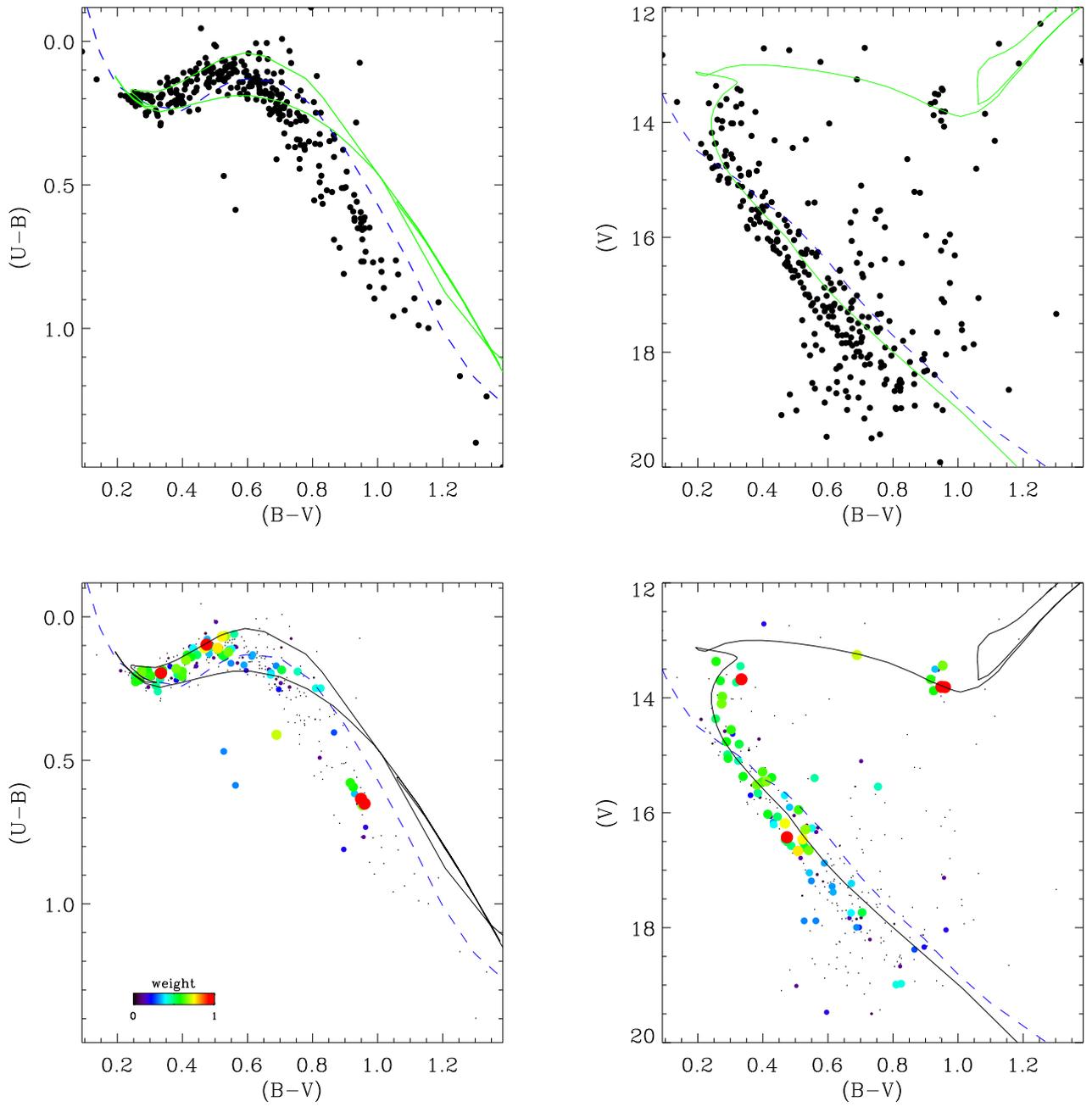}
\caption{Same as Fig.~\ref{fig3} for NGC~2266. }
\label{fig5}
\end{figure*}

\begin{figure*}
\centering
\includegraphics[width=18cm]{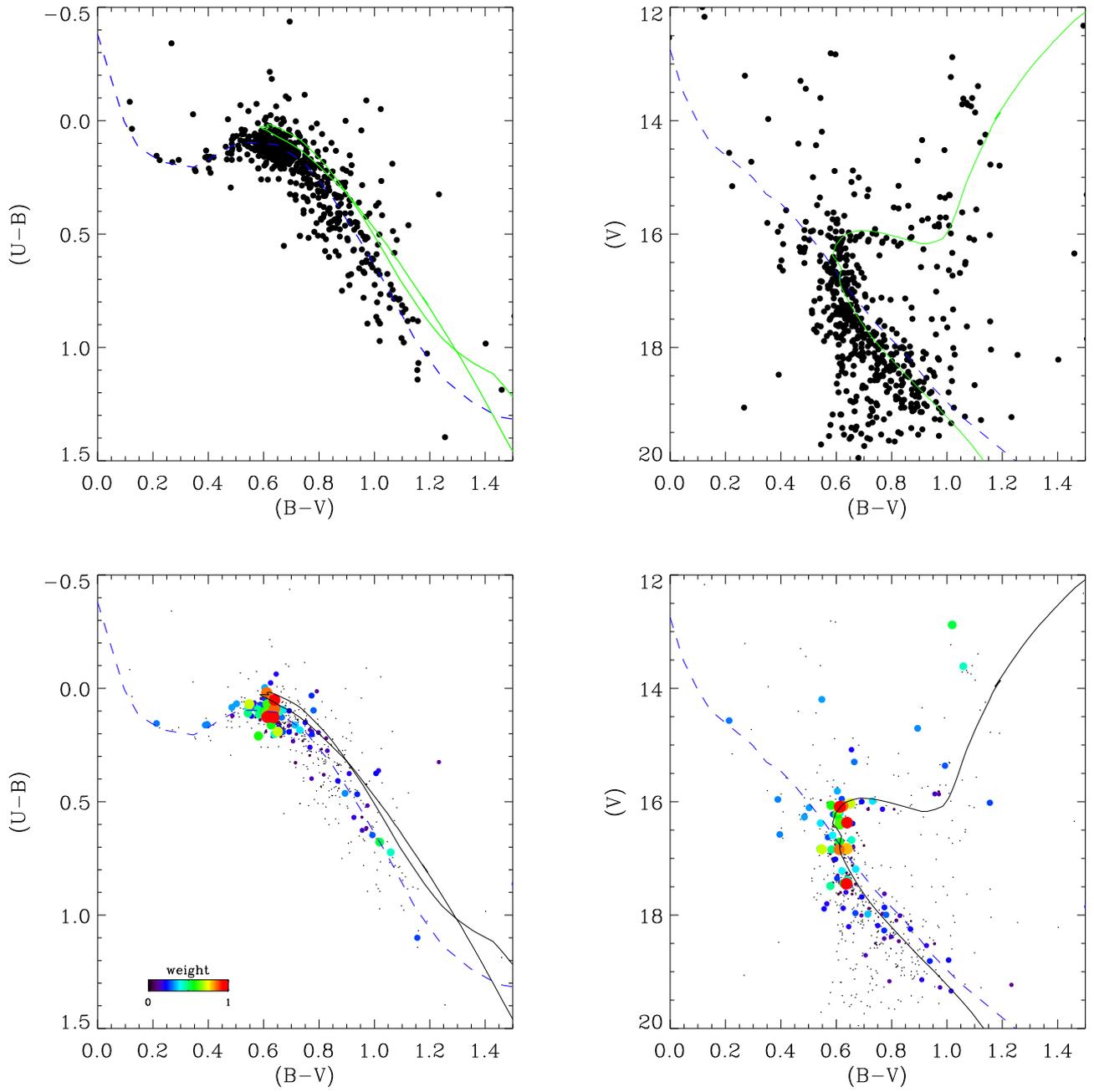}
\caption{Same as Fig.~\ref{fig3} for Berkeley~32. }
\label{fig6}
\end{figure*}

\begin{figure*}
\centering
\includegraphics[width=18cm]{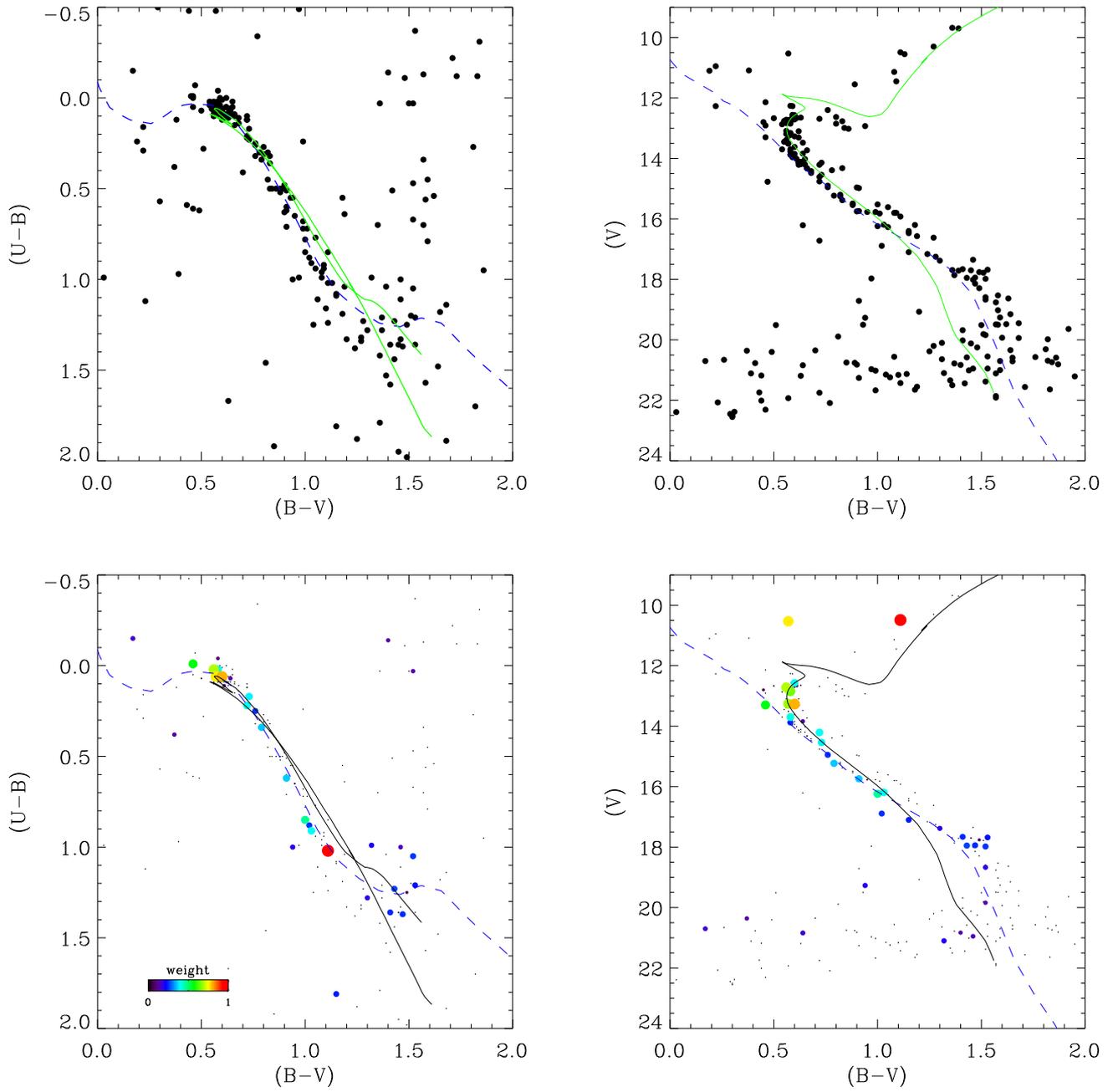}
\caption{Same as Fig.~\ref{fig3} for NGC~2682~(Ref. 31). }
\label{fig7}
\end{figure*}
\clearpage
\begin{figure*}
\centering
\includegraphics[width=18cm]{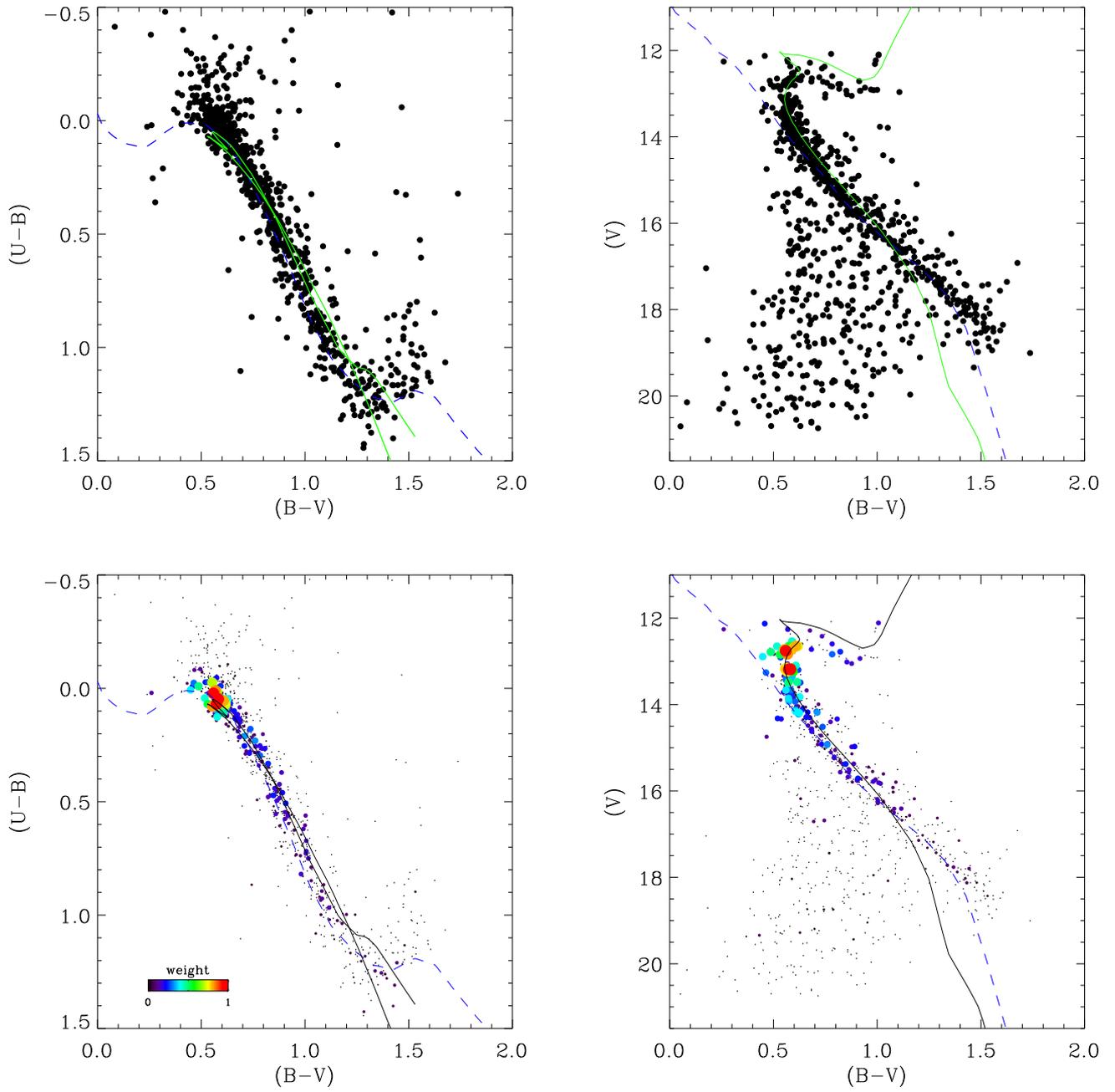}
\caption{Same as Fig.~\ref{fig3} for NGC~2682~(Ref. 54). }
\label{fig8}
\end{figure*}

\begin{figure*}
\centering
\includegraphics[width=18cm]{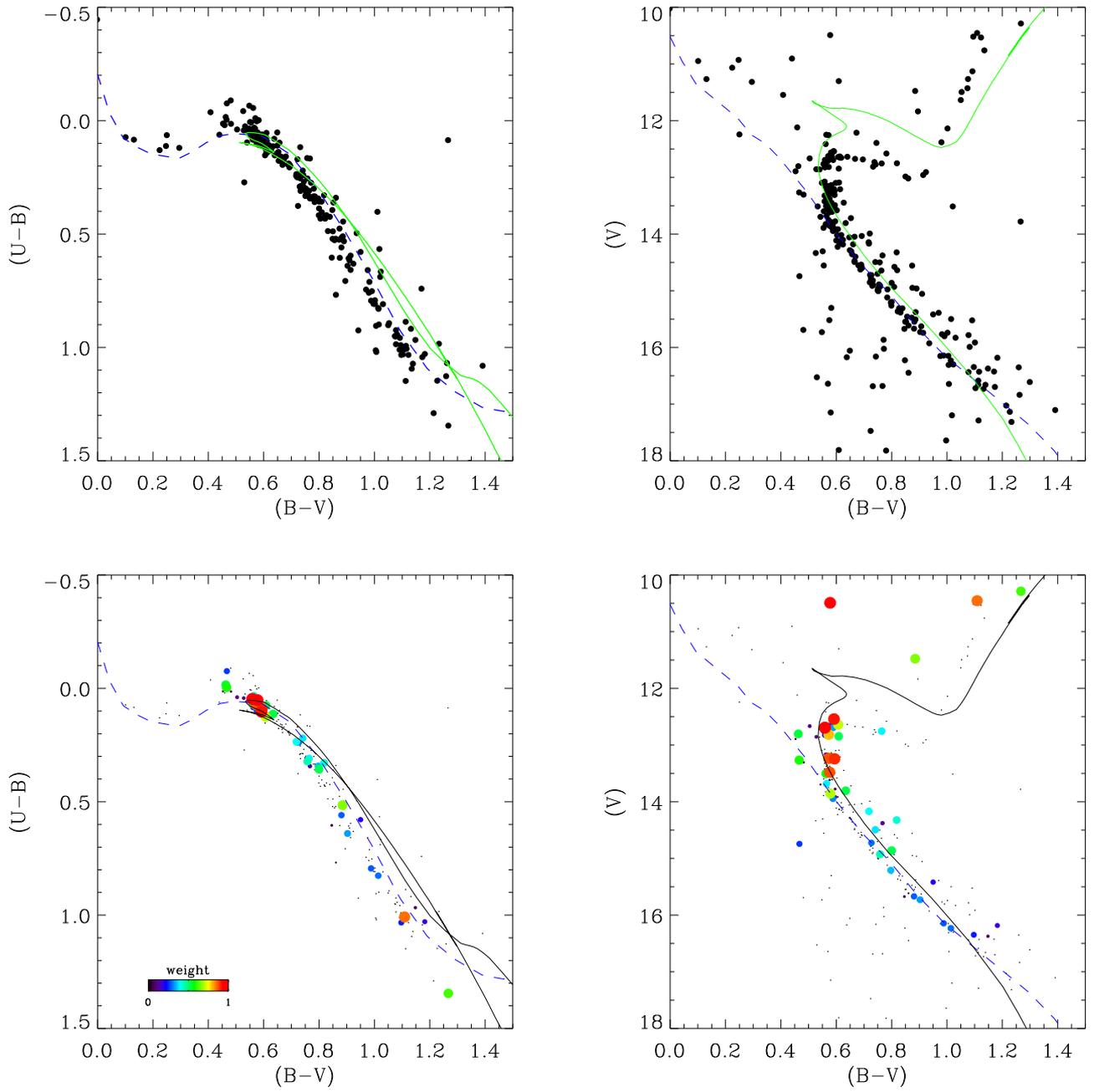}
\caption{Same as Fig.~\ref{fig3} for NGC~2682~(Ref. 335). }
\label{fig9}
\end{figure*}

\begin{figure*}
\centering
\includegraphics[width=18cm]{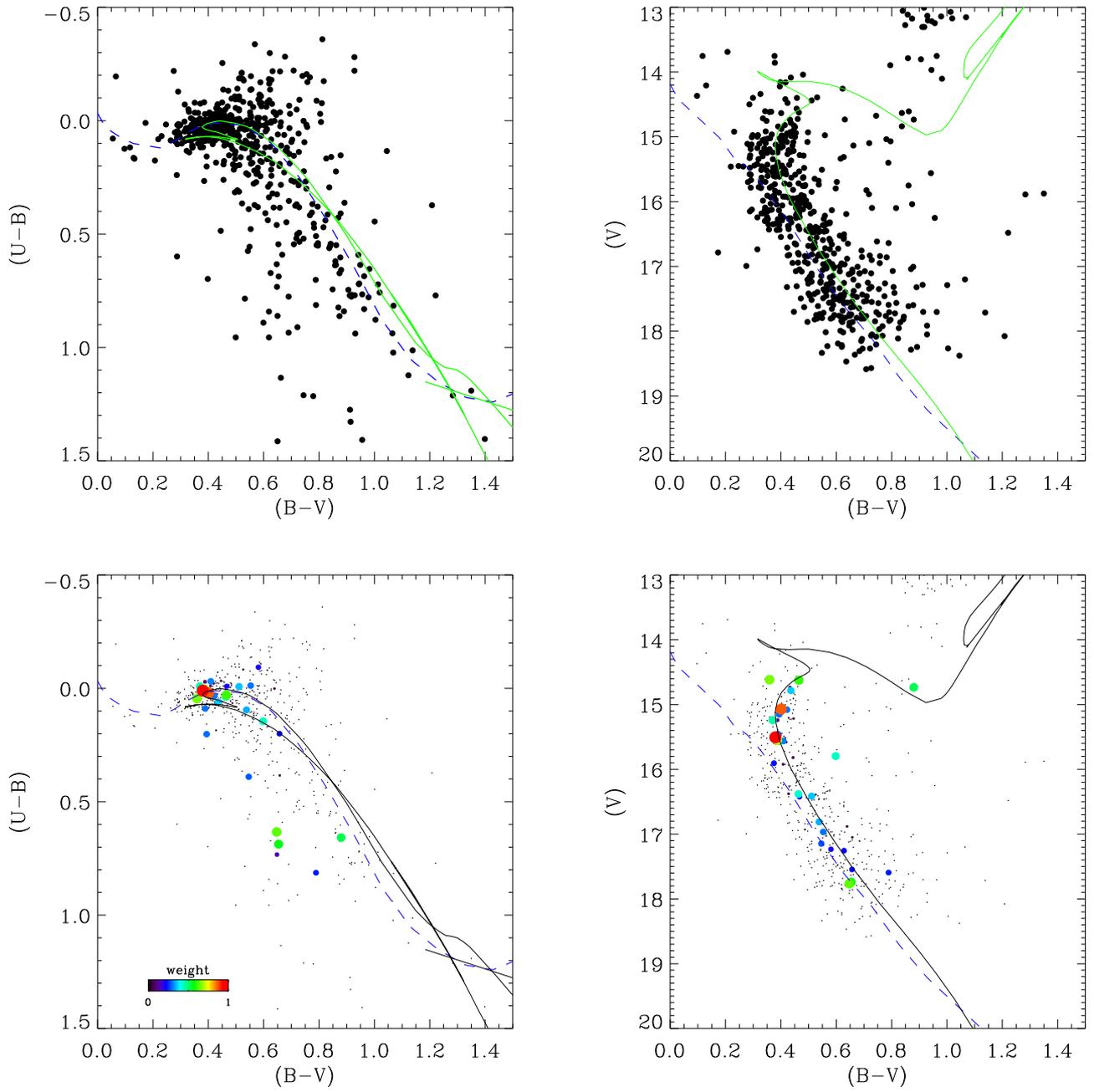}
\caption{Same as Fig.~\ref{fig3} for NGC~2506~(Ref. 284). }
\label{fig10}
\end{figure*}

\begin{figure*}
\centering
\includegraphics[width=18cm]{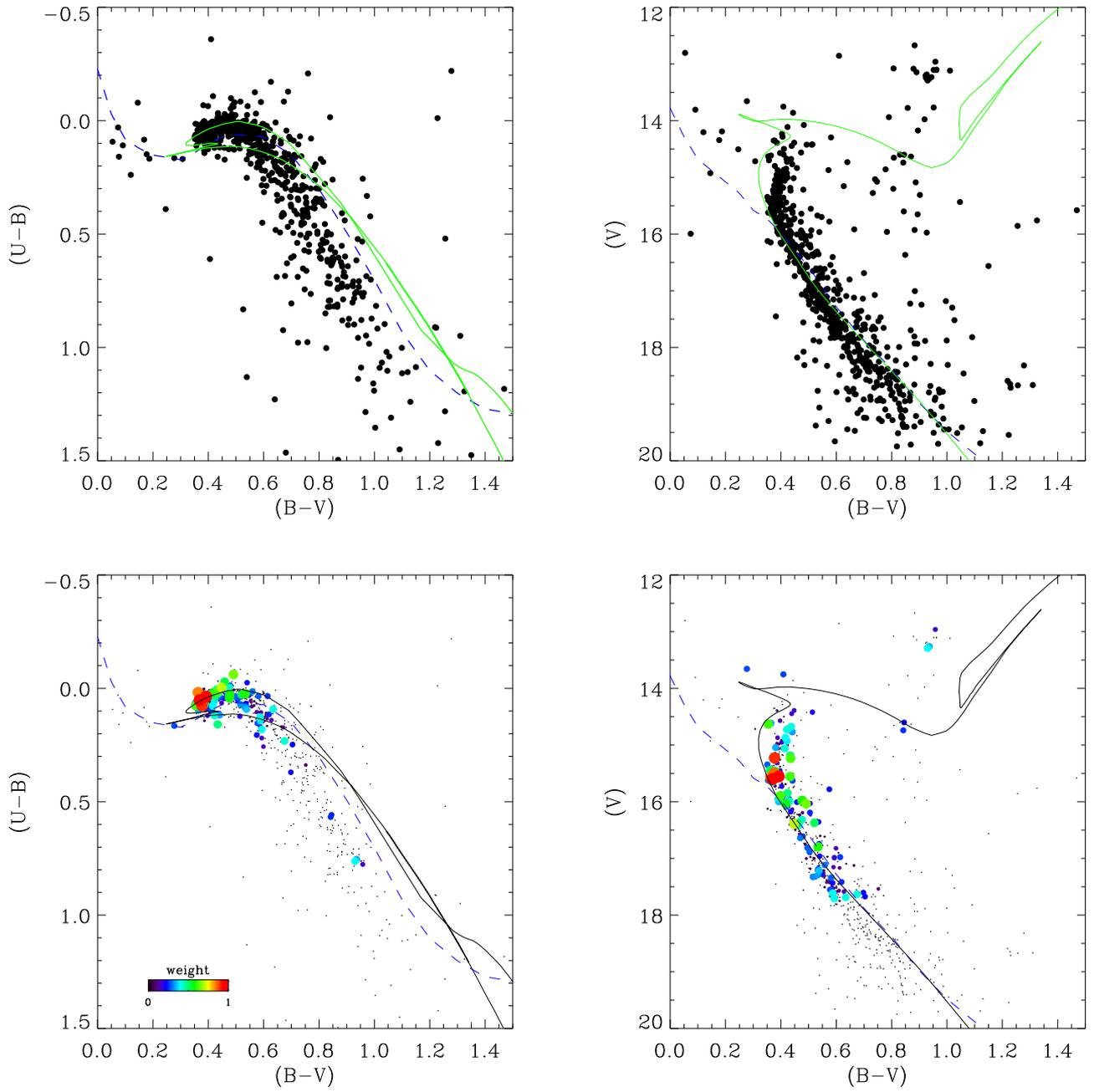}
\caption{Same as Fig.~\ref{fig3} for NGC~2506~(Ref. 163). }
\label{fig11}
\end{figure*}
\clearpage

\begin{figure*}
\centering
\includegraphics[width=18cm]{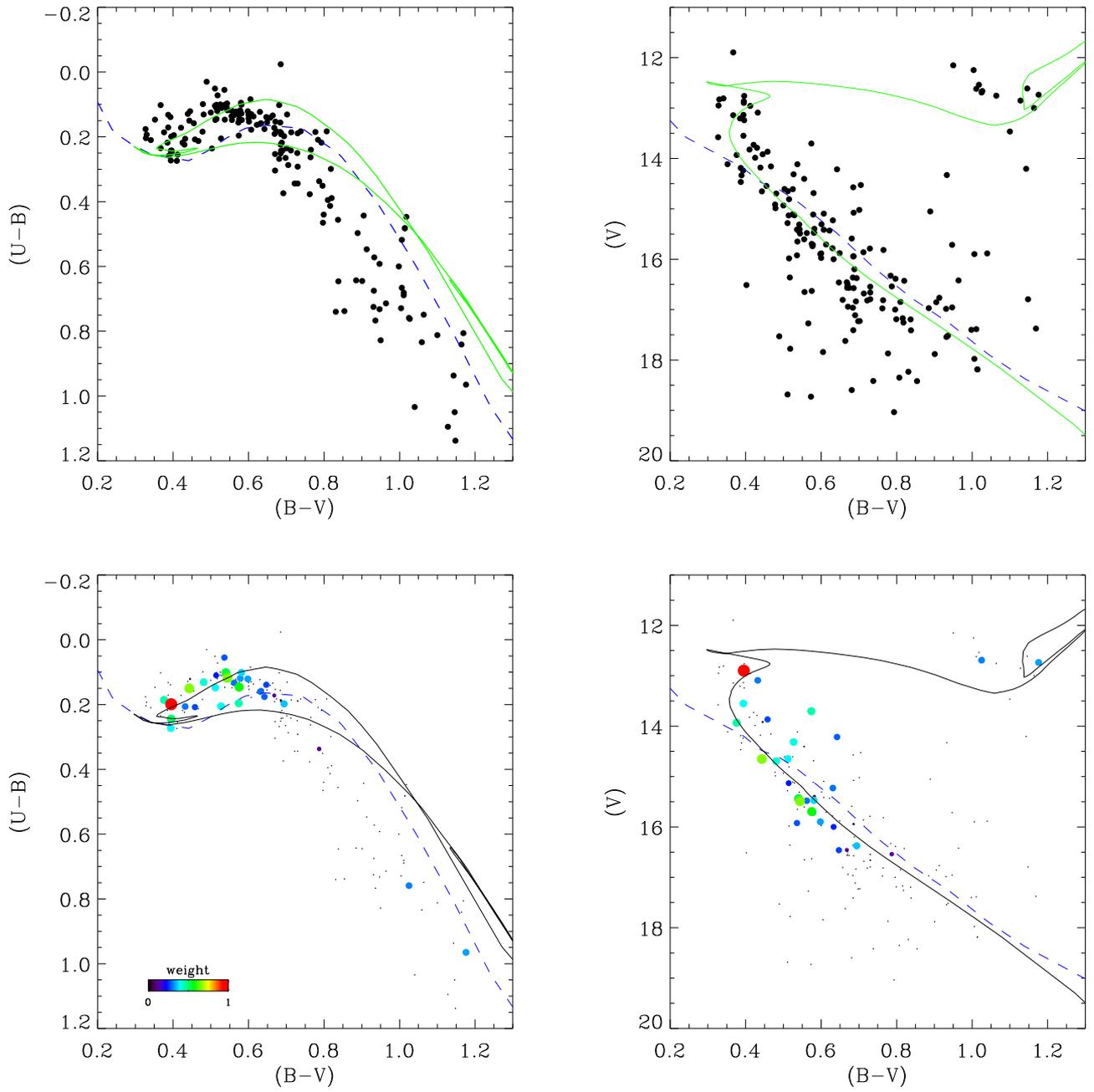}
\caption{Same as Fig.~\ref{fig3} for NGC~2355~(Ref. 44). }
\label{fig12}
\end{figure*}

\begin{figure*}
\centering
\includegraphics[width=18cm]{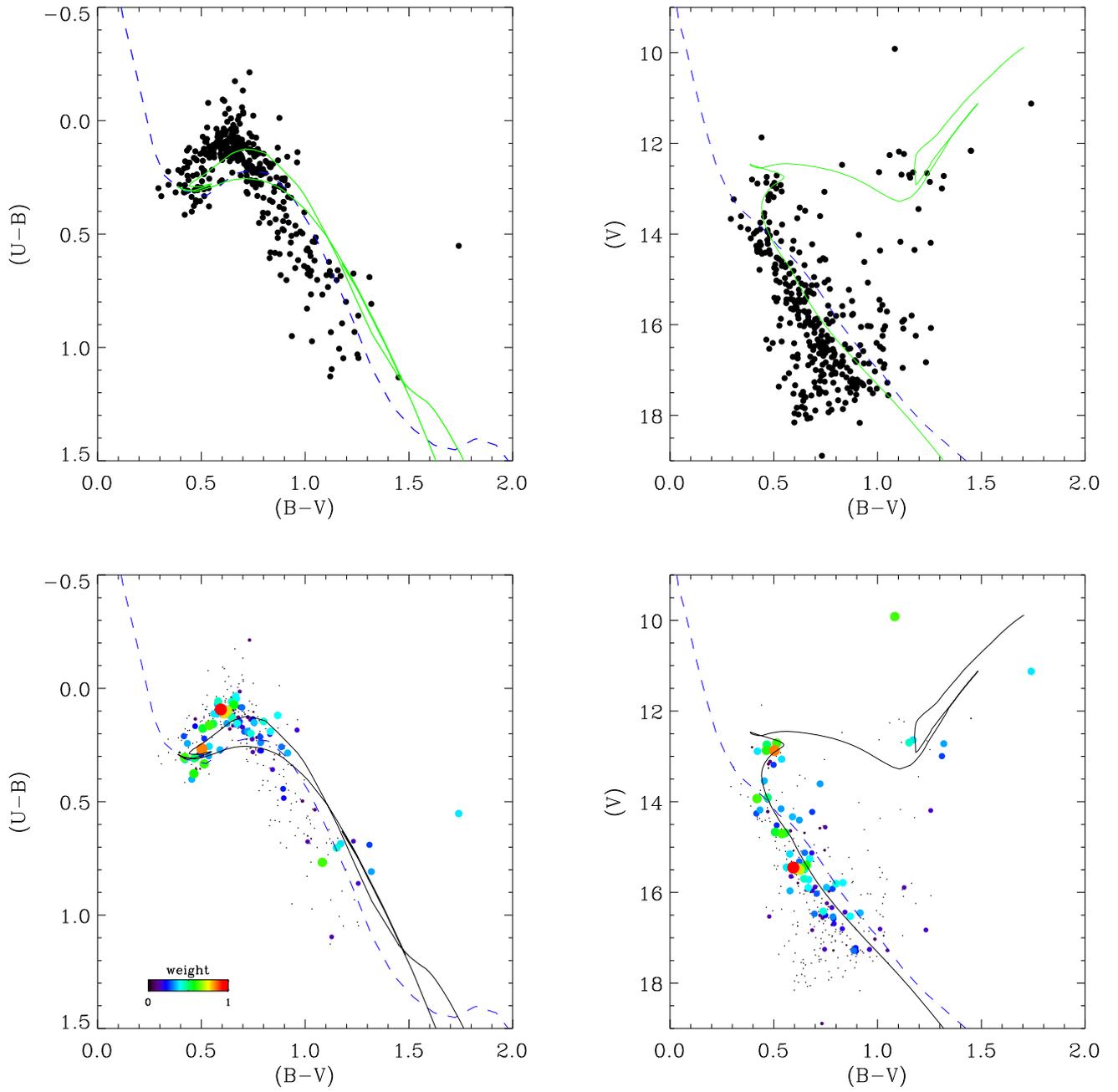}
\caption{Same as Fig.~\ref{fig3} for NGC~2355~(Ref. 217). }
\label{fig13}
\end{figure*}

\begin{figure*}
\centering
\includegraphics[width=18cm]{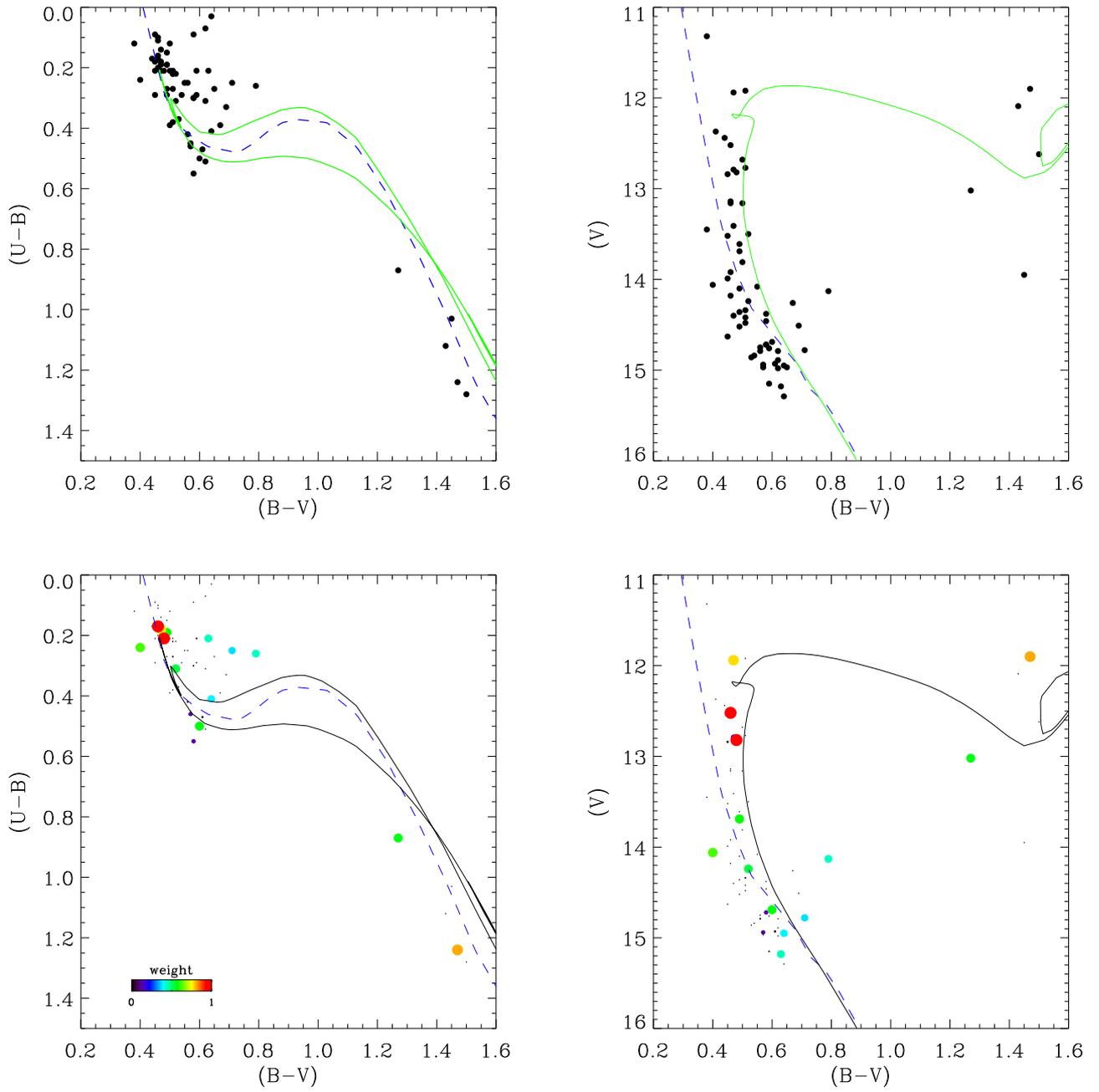}
\caption{Same as Fig.~\ref{fig3} for Melotte~105~(Ref. 289). }
\label{fig14}
\end{figure*}

\begin{figure*}
\centering
\includegraphics[width=18cm]{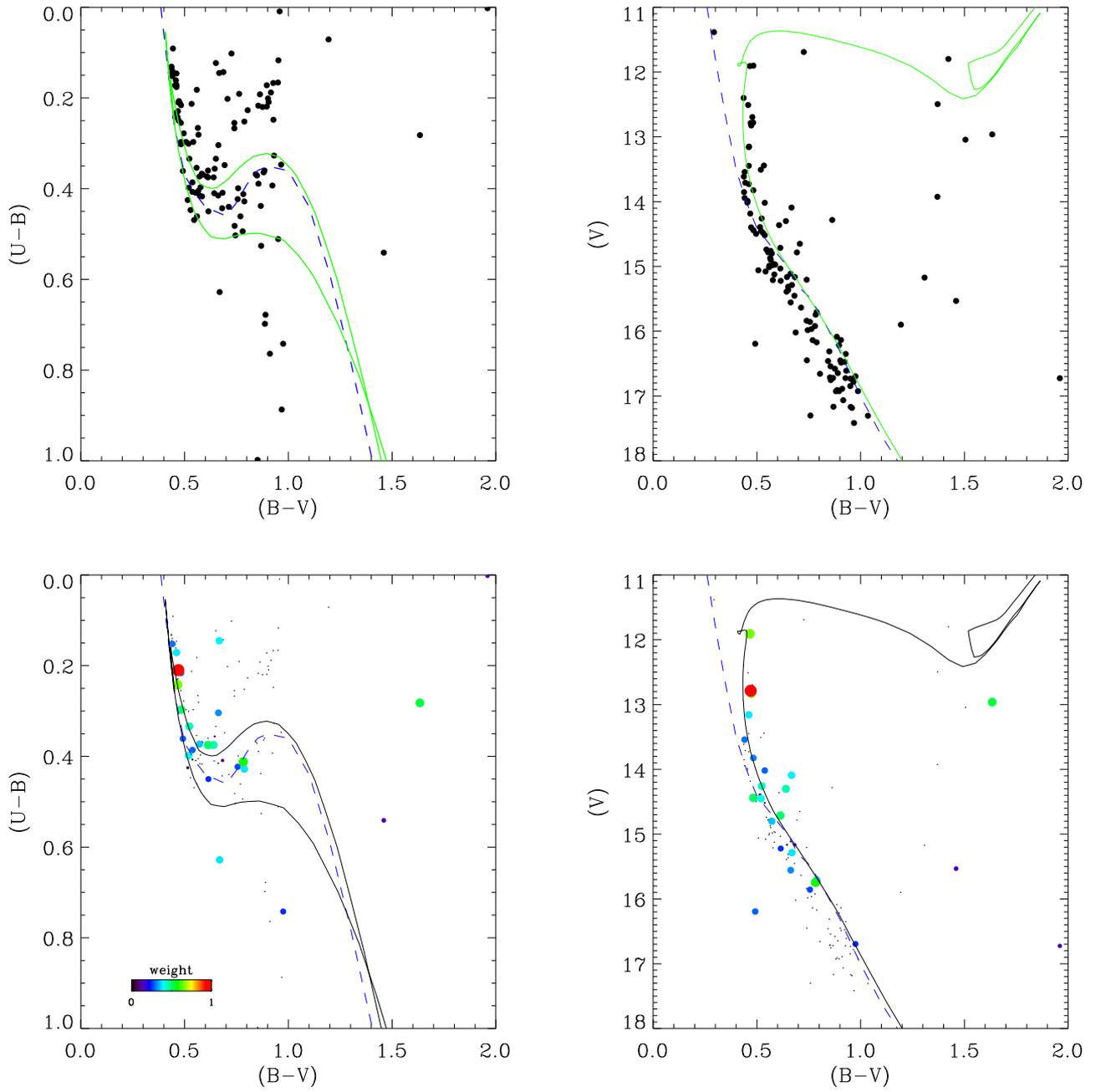}
\caption{Same as Fig.~\ref{fig3} for Melotte~105~(Ref. 32). }
\label{fig15}
\end{figure*}

\begin{figure*}
\centering
\includegraphics[width=18cm]{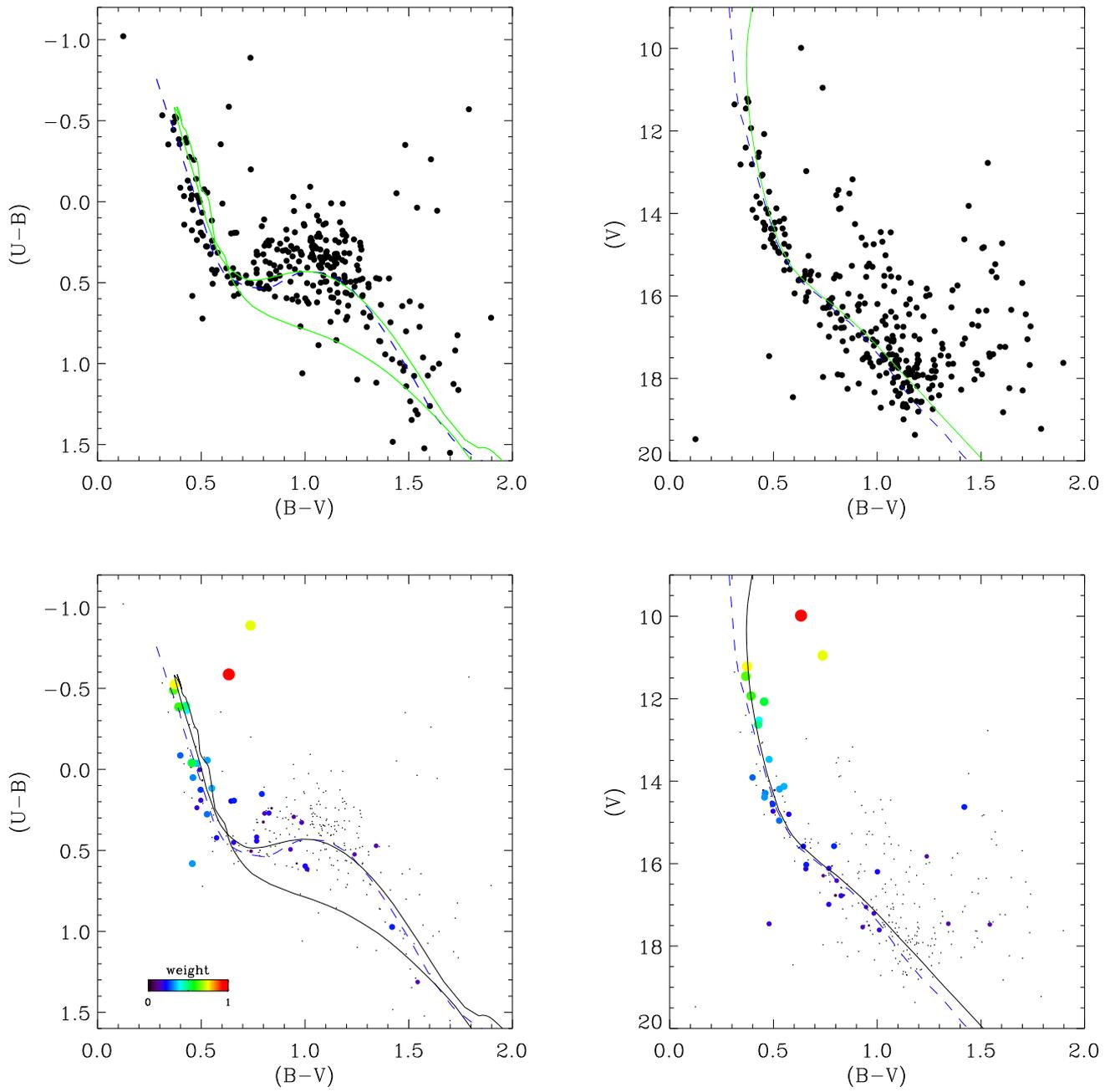}
\caption{Same as Fig.~\ref{fig3} for Trumpler~1~(Ref. 86). }
\label{fig16}
\end{figure*}
\clearpage
\begin{figure*}
\centering
\includegraphics[width=18cm]{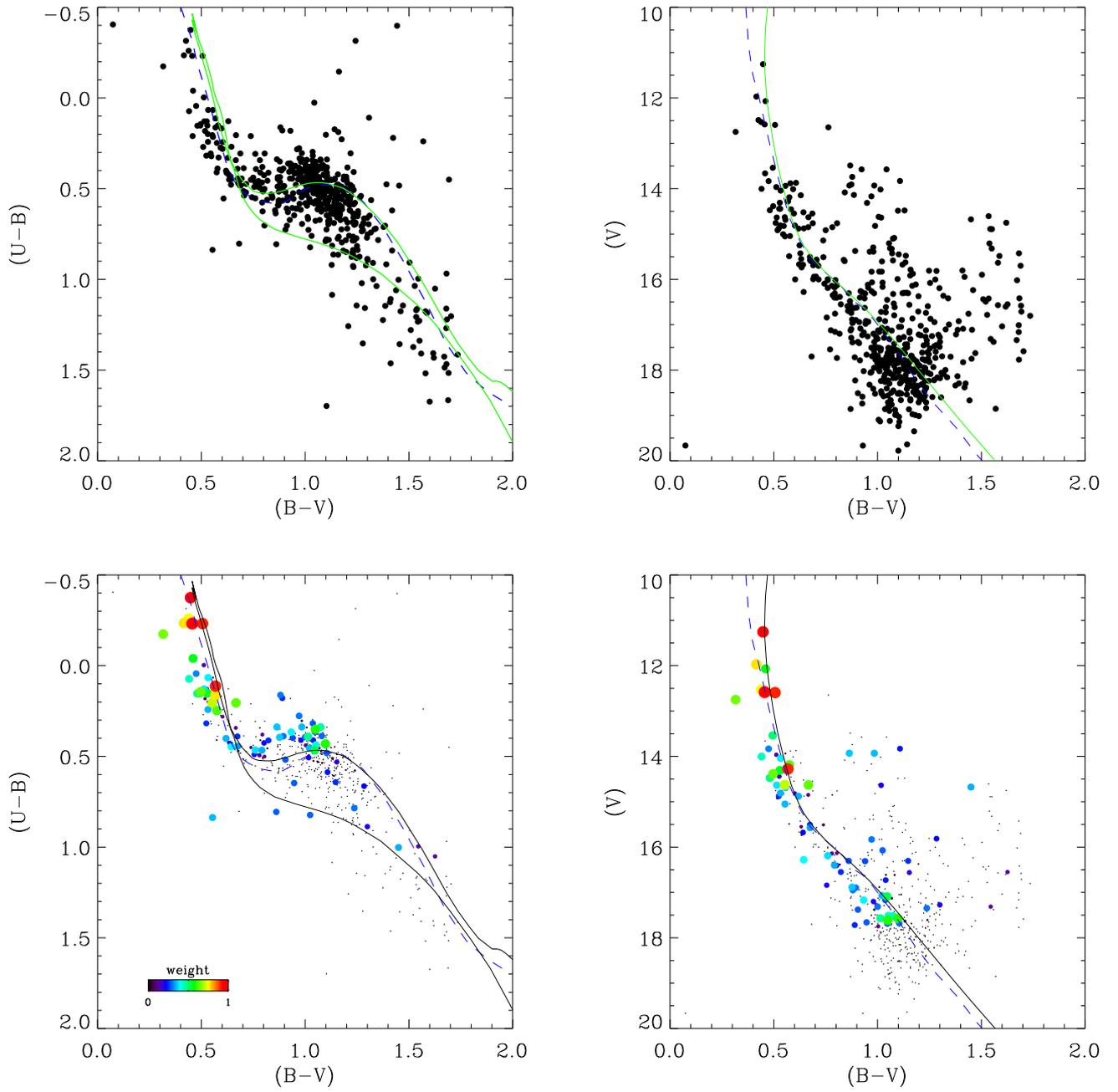}
\caption{Same as Fig.~\ref{fig3} for Trumpler~1~(Ref. 320). }
\label{fig17}
\end{figure*}
\end{appendix}

\end{document}